# Thermodynamic model of solute site preferences in ordered alloys

Gary S. Collins* and Matthew O. Zacate

Department of Physics, Washington State University, Pullman, WA 99164-2814

*email: collins@wsu.edu

## Abstract

A thermodynamic model based on the law of mass action is used to calculate concentrations of elementary point defects and to determine site preferences of solute atoms in ordered alloys. Combinations of lattice vacancies, antisite atoms and host interstitials that form equilibrium defects are enumerated for the CsCl (B2) and $Ni_2Al_3$ structures. For CsCl, in addition to the two substitutional sites, a distorted tetrahedral interstitial site is considered. For $Ni_2Al_3$, the Ni site, two distinct Al sites and a vacant Ni-type site that is operationally equivalent to an interstitial site are considered. Key to the model is the derivation of an equation of constraint among concentrations of elementary defects, for which an expression is given that is valid for any crystal structure. The concentration of a selected defect can be solved for using the equation of constraint in conjunction with mass-action equations describing formation of defect combinations. The method leads directly to defect concentrations without the need to evaluate composition-dependent chemical potentials, resulting in a transparent formalism in which all energy parameters are independent of composition and temperature.

The model is used to explore the phenomenology of site-preferences of dilute ternary solute atoms. A unified treatment is provided for interstitial and substitutional site preferences. The findings are in agreement with previous treatments, which were restricted to substitutional sites. Explicit expressions are worked out for site preferences of dilute solutes in CsCl and $Ni_2Al_3$. It is also shown how the model can be applied to other crystal structures and/or to systems in which concentrations of solute are not negligible in comparison with defect concentrations. General rules for how site-preferences depend on temperature and on composition in non-stoichiometric compounds are obtained through algebraic analysis and numerical simulations: (1) Solute S tends to occupy substitutional sites of element B in B-deficient compounds or of element A in A-deficient compounds. (2) If the difference of energies of S on sites A and B is very positive or negative, then S will occupy site B or A exclusively, independent of composition. If the difference of site energies is intermediate, the solute will switch from one site to the other as the composition changes. (3) Solutes have a tendency to occupy interstitial or empty-lattice sites with a maximum site-fraction near the stoichiometric composition.

I.  INTRODUCTION

There is considerable interest in site preferences of solutes in compounds. A solute can have an important influence on material properties in concentrations of the order of an atomic percent. Site preferences of solutes in NiAl, for example, have been correlated with various changes in properties.[1] An important issue is to interpret observed site preferences in terms of underlying atomic interactions.

Experimental methods that have been used to measure site preferences include x-ray diffraction and neutron diffraction, ALCHEMI,[2] analysis of solubility lobes in ternary phase diagrams,[3] and ridge detection of thermal conductivity.[4] These methods, however, have mostly been applied to compounds of simple structure such as CsCl (B2) and $Cu_3Au$ ($L1_2$), in which all substitutional sites for each element are equivalent. Moreover, the methods are macroscopic and typically require solute concentrations of the order of a percent in order to detect site preferences. At such high concentrations, site preferences may differ from those that would be observed in the dilute limit.

As an alternative to the above methods, one can identify site preferences microscopically through measurement of hyperfine interactions of probe nuclei. Hyperfine interactions depend on local atomic environments of the probes and thus, in principle, can be used to identify lattice locations of solute atoms. One then has an atom-scale 'fingerprinting' technique to determine site preferences of the probes. Using this idea, Mössbauer spectroscopy and nuclear magnetic resonance have been used to determine site preferences of ternary solutes in $Fe_3Si$ through measurements at host probes.[5] A second approach is to identify lattice locations of impurity probes (as opposed to host probes) through their hyperfine interactions. This idea has been applied, for example, in a nuclear quadrupole resonance study of site preferences of Cu solutes in $Ni_2Al_3$ (ref. [6].)

Perturbed angular correlation of gamma rays (PAC) has been applied extensively to study local environments of probes in solids. In particular, point defects next to impurity probe atoms have been detected through disturbances to hyperfine interactions in metals and intermetallics[7,8] and in semiconductors.[9] Implicit in most of these studies has been an assumption about the lattice location of the probe atom. Efforts to determine lattice locations of probes through measurement of hyperfine interactions have been few.[10,11,12] In addition, interstitial sites have generally not been explicitly considered because of assumptions that they would have very high site-energies when impurities and host atoms have similar atomic radii. An early effort was made to determine the site-preference of 1 at.% of Hf solutes in $Ni_3Al$, which has the $Cu_3Au$ structure. The basic idea is as follows. The Al-site has cubic point symmetry so that there should be no efg and zero quadrupole interaction frequency for a probe at that site, whereas the Ni-site has a four-fold axis of charge symmetry and the quadrupole interaction should be non-zero at that site. In other situations, sites can be difficult to distinguish in this way; for example the two cubic substitutional sites in the perfect CsCl structure would both exhibit zero quadrupole interactions. However, if there is a wide phase field and if the structural point defects are known, one can infer the lattice site of an impurity probe through strengths of efgs produced by the defects. In particular, one can usually identify if defects are in the nearest atomic shell or farther away. Extensive PAC studies have been made in this laboratory of point defects next to In probe atoms in intermetallics having the CsCl structure, including NiAl, CoAl,

FeAl, NiGa, CoGa (ref. 13.) For these systems, strengths of hyperfine interactions of known TM-vacancies indicate that In solutes always site on the sublattice of the trivalent element, at least in alloys rich in the transition-metal (TM). For TM-poor alloys, information is more fragmentary. For Ni-poor NiAl, indium solutes clearly site on the trivalent sublattice, with large-frequency interactions observed for Ni-vacancies in the first shell,[14] but in FeAl (ref. 15) and NiGa (ref. 16) the situation has been less clear. For more complex crystal structures, knowledge of the lattice location of an impurity probe atom is more often than not guesswork.

In this situation, it is desirable to have a model that both gives insight into the phenomenology of site-preferences and allows for detailed calculations. At the most naïve level, the ratio of site-fractions of solutes on two different sites α and β in a compound might be expected to vary as $f_\alpha / f_\beta \approx \exp(-(G_\alpha - G_\beta)/k_B T)$, in which $(G_\alpha - G_\beta)$ is the difference of energies of the solute atom in the two sites. Since site-energies are determined by the local atomic configurations, site-fraction ratios would be expected to be largely independent of composition for small deviations from stoichiometry. However, it has been shown that concentrations of solute on substitutional sublattices are coupled to concentrations of intrinsic defects through reactions that transfer solutes between different sites and help to establish an equilibrium distribution of solute.[17, 18, 19] The present model treats the more general situation in which defects and solutes may occupy substitutional and interstitial solute sites and in which there may be inequivalent sublattices of an element of the compound.

Recent PAC experiments in this laboratory, to be published separately, have revealed a strong and interesting correlation of the location of indium solutes with composition in $Ni_2Al_3$ phases.[20] In five phases, the dominant solute location was found to be different on the two sides of the stoichiometric composition, suggesting as above a correlation with concentrations of structural point defects. The $Ni_2Al_3$ structure is more complex than CsCl, with one TM-type site (designated α), two inequivalent trivalent-metal sites (β1 and β2) and one vacant TM-type site (X). Only $10^{11}$ $^{111}$In probes are required for a measurement, so that the typical fractional solute concentration was 10 parts-per-billion in a 100-mg sample. Such concentrations are well below concentrations of relevant intrinsic defects, ensuring that observed site preferences are independent of the solute concentration. For galliumides $Ni_2Ga_3$ and $Pt_2Ga_3$ and aluminides $Ni_2Al_3$, $Pd_2Al_3$ and $Pt_2Al_3$ it was found that In occupies site β2 for TM-rich compositions. For Al-rich aluminides, it was found that indium is forced to ill-defined non-crystallographic sites such as in grain boundaries. However, for Ga-rich galliumides it was found that indium exhibits two signals with low efg's, one prominent near stoichiometry and the other dominant in more Ga-rich samples. From the crystal structure, described below, it can be shown that there are two sites with low efg's: the Ni-site α and the empty-lattice site X. As far as concerns defect chemistry, empty-lattice sites are equivalent to interstitial sites, although they may have larger volumes and therefore more readily be occupied by solute atoms of the same size as host atoms. These observations motivated development of a thermodynamic model of site preferences that could describe the phenomenology of site preferences on both substitutional and interstitial-like sites.

Thermodynamic models of defects in intermetallic compounds have most often been formulated starting from the free energy of the compound.[18, 21, 22, 23, 24, 25, 26, 27, 28, 29, 30, 31] Usually, the free energy is minimized with respect to defect concentrations in order to obtain expressions for defect concentrations either as functions of defect enthalpies and chemical potentials [18, 21, 24, 25, 29, 30] or as functions of effective defect formation enthalpies.[18, 21, 24] A number of ways has been used to handle the chemical potentials and effective formation enthalpies, which are functions of composition: (1) numerical solution of coupled non-linear equations,[21,29] (2) simplification under the assumption of particular structural defects,[18,23,24,25,30] or (3) further manipulation of equations to eliminate chemical potential terms in defect concentrations by inclusion of explicit composition dependence.[26, 28, 30] An approach related to starting from the free energy is to start from the grand partition function.[32] In all the above approaches, assumptions are made that the concentrations of defects are dilute and that the defects are non-interacting at stages at which specific results are calculated.

The approach used in this study instead is to start from a general equation of constraint that contains all structural and compositional information and to apply the law of mass action for possible defect reactions. This is similar to the approach used in reference 17. Use of the law of mass action also has been suggested as a starting point for thermodynamic derivation of defect concentrations in non-stoichiometric compounds.[25,33] Application of the law is restricted to non-interacting defects and low defect concentrations. Its use is therefore appropriate for intermetallic compounds in which interactions between defects are short-ranged due to screening by the conduction electrons and also for other systems such as insulators containing isovalent impurities. Assuming that defects are non-interacting and their concentrations are dilute, equations for equilibrium constants obtained via the law of mass action are rigorous and can be obtained through minimization of the crystal free energy subject to structural and compositional constraints.[22, 34] For example, Ren and Otsuka derived mass-action equations for defects in the CsCl structure starting from the crystal free energy using a bond model. Similar to the present work, they used a compositional constraint in conjunction with mass-action equations to determine defect concentrations. The present work illustrates more generally how to calculate concentrations of intrinsic defects and solutes in a binary compound of arbitrary structure using a structural and compositional constraint and mass-action equations. Specific examples are given for CsCl and $Ni_2Al_3$ structures. A framework is then established within which one can calculate the partition of solute atoms among substitutional and interstitial sites. The present work reproduces prior results for substitutional sites,[17, 18, 19, 21, 35, 36] as discussed in Section VI, and goes further by treating substitutional and interstitial sites in a unified way.

The organization of the paper is as follows. The model is applied first for the simple, common and technologically important CsCl structure and then for the more complex, multi-site $Ni_2Al_3$ structure. The law of mass action is used to obtain relations among concentrations of elementary intrinsic defects and to determine ratios of fractions of solutes on different sites. For these structures, solutes are assumed to be present in infinite dilution, as is appropriate for interpretation of the PAC experiments. For CsCl an interstitial site is considered in addition to the two substitutional sites. The defect chemistry of $Ni_2Al_3$ is examined completely. The structure has a number of interesting features, including two inequivalent aluminum sublattices and a vacant sublattice that can be populated by atoms jumping to it from occupied sublattices. The vacant sublattice has the character of the interstitial site considered in the CsCl structure. Numerical simulations are made to explore qualitative features of the model and to identify phenomenological rules for solute site preferences. In particular, the dependence of site preferences on deviations from stoichiometric compositions is examined. It will be shown, as in previous work restricted to consideration of substitutional sites,[18] that each site-fraction ratio is proportional to (1) an intrinsic defect concentration and to (2) a thermally-activated factor involving the sum of the site-energy difference and, as well, the energy of the same intrinsic defect. Thus, a site-fraction ratio varies with composition in the same way as an intrinsic defect. A general approach applicable for a binary compound of arbitrary structure and for non-negligible concentrations of a solute is shown in an appendix. Previous models and calculations of site-preference behavior are reviewed along with consideration of the influence of the width of the phase field on the observability of changes in site-preference.

## II. DEFECTS IN AB COMPOUNDS

The thermodynamics of substitutional and interstitial defects in binary compounds that are stoichiometric at the 1:1 atomic ratio of elements is considered in this section. The analysis is described in terms of the CsCl structure, with one sublattice for each element and one unique type of interstitial site, but is equally applicable for other 1:1 structures such as NaCl (B1) and CuAu (L1$_0$). The structure of CsCl is shown in Fig. 1. It consists of two interpenetrating substitutional sublattices $\alpha$ and $\beta$, normally occupied by A and B atoms, respectively. Elementary intrinsic point defects considered are vacancies, $V_\alpha$ and $V_\beta$, and antisite atoms, $A_\beta$ and $B_\alpha$, on the substitutional sublattices and atoms $A_\tau$ and $B_\tau$ on interstitial sites $\tau$. Site $\tau$ is defined as the distorted tetrahedral-interstitial site located at permutations of positions ($\pm 1/2$, 0, $\pm y$), in which y is in the range 0 and 1/2. There are 24 such $\tau$–sites on the surface of the conventional cubic cell, making 12 equivalent sites per cell. Four of these are indicated for y= 1/4 in the figure. For A and B atoms having typical radii of order 0.15 nm, such an interstitial site would tend to admit atoms of small radius, e.g. boron. We will assume that host or solute interstitial atoms are large enough to block the occupation of neighboring $\tau$–sites by additional interstitials, so that only a maximum of 3 $\tau$-sites can be occupied per substitutional site. In the limit y=0, $\tau$-sites are at centers of the six cube faces (at one type of distorted octahedral interstitial site) whereas for y=1/2 they are at the middles of the twelve cube edges (at a second type of distorted octahedral interstitial site.)

*(a) Equation of constraint among defect concentrations*. Many phases have an extended field width about the stoichiometric composition. A classic example is NiAl, which maintains the CsCl structure for compositions between 45 and 60 at.% Ni. Even 'line' compounds must have phase fields of finite, albeit narrow, width. Broad phase fields are made possible by incorporation of structural and/or thermal point defects. For any specified binary crystal structure, an equation of constraint among defect concentrations exists that is a consequence of the fact that proportions of elements in the compound are fixed. A rigorous method for constructing the equation is given in the Appendix. Consider a generic binary compound $A_{1+2x}B_{1-2x}$, in which the deviation from the stoichiometric 1:1 composition is measured by parameter $x$ (e.g., $x$=+0.01 corresponds to a sample with 51 at. % of element A.) The solute concentration is assumed to be negligible in comparison with concentrations of intrinsic defects. Using the method in the Appendix, fractional concentrations of the six elementary defects are found to be constrained by the following relation:

$$2[B_\alpha] + (1-2x)[V_\alpha] + 3(1+2x)[B_\tau] + 4x = 2[A_\beta] + (1+2x)[V_\beta] + 3(1-2x)[A_\tau]. \quad (1)$$

This equation and corresponding constraint equations for other structures are key to the application of the law of mass action to solve for defect concentrations. Square brackets indicate fractional concentrations of defects, with the sublattice on which the defect sits indicated by the subscript. The fractional concentration is defined as the number of defects divided by the total number of sites on the sublattice. This leads to simpler expressions than when defined with respect to all lattice sites (e.g., in ref. 22) or all atoms (e.g., in ref. 23) since the ranges of concentration have natural limits of 0 and 1.

Defects that appear on the left-hand side of eq. 1 can be compensated in different ways by defects on the right-hand side. For example, at the stoichiometric composition ($x=0$) the two antisite defect concentrations must be equal if other defect concentrations are zero, and similarly for vacancy or interstitial defects. Alternatively, two $V_\alpha$ defects on the left-hand side can balance one $A_\beta$ defect on the right-hand side in good approximation (the so-called triple defect[37].) A deviation from stoichiometry by an amount $x=-0.01$ could in principle be accommodated by structural defects in concentrations of either $[V_\alpha] \cong 0.039$, $[B_\alpha] = 0.020$ or $[B_\tau] \cong 0.014$ alone. Note that three possible structural defects in A-poor compounds ($V_\alpha$, $B_\alpha$, $B_\tau$) and three opposite types in A-rich compounds ($V_\beta$, $A_\beta$, $A_\tau$) are grouped on left- and right-hand sides of eq. 1.

*(b) Equilibrium defect combinations.* Thermally activated defects can only occur in combinations of opposing structural defects that appear in equation 1. There are six elementary defects, and nine possible paired combinations of opposing structural defects. Of the nine pairs, five are sufficient to establish couplings among all defect concentrations and then, using the law of mass action, to solve for defect concentrations. These are chosen as follows. (i) The antisite atom pair, formed by interchanging atoms on the two sublattices ($A_\beta + B_\alpha$). (ii) The Schottky vacancy pair, which can be thought of as being formed by removal of a molecule of the compound from the interior of the crystal with placement on the surface, thereby increasing the volume of the crystal by one unit cell and leaving behind one vacancy on each sublattice ($V_\alpha + V_\beta$). (iii) The triple-defect comprised of two A-vacancies ($V_\alpha$) and one antisite A atom on the B sublattice ($A_\beta$), i.e., ($2V_\alpha + A_\beta$). The triple defect can be formed from a Schottky pair by transfer of an $A_\alpha$ atom into a B-vacancy, creating an $A_\beta$ antisite atom and second A-vacancy: $V_\alpha + V_\beta + A_\alpha \to 2V_\alpha + A_\beta$. (iv) A Frenkel defect involving an A-interstitial, called Frenkel-A defect below, formed by transfer of an A-atom from the $\alpha$- to $\tau$-sublattice, leaving a vacancy on the $\alpha$-sublattice ($V_\alpha + A_\tau$), and (v) a Frenkel-B defect ($V_\beta + B_\tau$). The crystal volume is thus increased by one unit cell upon formation of a Schottky or triple defect, but not upon formation of an antisite pair or Frenkel defect. It will be uniformly assumed below that all defects and solutes are non-interacting and therefore randomly distributed on their sublattices.

Four other combinations can be formed out of the five chosen above. These are a second type of triple defect ($2V_\beta + B_\alpha$), an interstitial pair ($A_\tau + B_\tau$), and two combinations of antisite defects with interstitials ($A_\beta + 2B_\tau$) and ($B_\alpha + 2A_\tau$). The interstitial pair can arise, for example, by formation of a Frenkel-A defect and of a Frenkel-B defect, followed by annihilation of a Schottky vacancy pair. Finally, there is no need to consider formation of combinations consisting of more than two different defects because they can be decomposed into combinations of just two types of defect.

Table I

The free energy of the crystal is minimized at elevated temperature by formation of point defect combinations. Formation reactions for the five 'fundamental' defect combinations described above are given in Table I. Also given in the table are expressions for equilibrium constants of the formation reactions in terms of products of fractional concentrations of defects. These expressions come from the law of mass action,[22,34] which for a reaction of stoichiometry $aA + bB \rightarrow cC + dD$ gives for the equilibrium constant $K = [C]^c[D]^d[A]^{-a}[B]^{-b}$. In the table, $0$ represents the perfect, defect-free lattice, $K$'s are equilibrium constants for the reactions and $G$'s are free energies of formation of defect combinations. The first identifying subscript of the free energies and equilibrium constants $G$ and $K$ in Table I is the number of elementary defects comprising the particular defect combination. To a good approximation, it can be shown that the dominant thermally activated defect combination will be the one having the smallest value of $G$ divided by the number of comprising elementary defects.

One can write more rigorous expressions for equilibrium constants. For example, the antisite pair forms via the reaction $A_\alpha + B_\beta \rightarrow A_\beta + B_\alpha$, for which the equilibrium constant is rigorously $K_{2A} = \frac{[A_\beta][B_\alpha]}{[A_\alpha][B_\beta]} = \frac{[A_\beta][B_\alpha]}{(1-[B_\alpha]-[V_\alpha])(1-[A_\beta]-[V_\beta])}$. Normally, concentrations of host atoms on normal sites, $[A_\alpha]$ and $[B_\beta]$, are assumed to be sufficiently close to unity in such expressions that they can be set to unity. While one can use the more rigorous expressions, for large defect concentrations the assumption of well-defined free-energies of formation of defects breaks down anyway due to the presence of neighboring defects. Therefore, we set the host-atom concentrations to unity in equations for equilibrium constants.

Values of $G$ for all nine defect combinations must be positive for the crystal structure to be stable. Free energies of the four defect combinations not listed in Table I can be written as sums and differences of $G$'s for the five listed combinations. For example, the free energy of formation of the interstitial pair ($A_\tau + B_\tau$) is equal to $G_{2FA} + G_{2FB} - G_{2V}$. Formation of some defect combinations is accompanied by an increase or decrease by one in the number of unit cells of the crystal. Such changes have to be accompanied, respectively, by an increase or decrease in the formation energy by an amount equal to the cohesive energy per unit cell. The number of cells increases by one for formation of a vacancy pair or triple-defect and decreases by one for formation of an interstitial pair or of one of the two combinations of antisite atoms and interstitials described above. In Table I, increases in the number of unit cells are indicated explicitly by the notation +**uc**.

*(c) Defect transformations.* Formation reactions in Table I can be used to generate additional reactions in which defects transform among themselves. For this purpose, a defect on one side of a reaction can be replaced by its *antidefect* on the other side. This device leads to useful relations especially for complex structures. For the CsCl structure, the defect-antidefect pairs are $(A_\beta, B_\alpha)$, $(V_\alpha, A_\tau)$, $(V_\beta, B_\tau)$. Thus, for example, line 3 in Table I implies a transformation reaction among equivalent structural defects: $B_\alpha \rightarrow 2V_\alpha$ +uc. Similarly, +uc on one side of a reaction can be replaced by -uc on the other so that, e.g., $2V_\alpha \rightarrow B_\alpha$ -uc. Finally, the antidefect of triple-defect combination $2V_\alpha + A_\beta$ is $B_\alpha + 2A_\tau$.

*(d) Solving for equilibrium defect concentrations.* Eq. 1 can be rewritten in terms of a single defect concentration by eliminating all other defect concentrations using expressions for the five equilibrium constants in Table I. Expressed in this way in terms of $[V_\alpha]$, for example, eq. 1 becomes a quartic polynomial equation:

$$2\frac{K_{2A}}{K_3}[V_\alpha]^4 + \left\{(1-2x) + 3(1+2x)\frac{K_{2FB}}{K_{2V}}\right\}[V_\alpha]^3 + 4x[V_\alpha]^2 - \\ - \left\{(1+2x)K_{2V} + 3(1-2x)K_{2FA}\right\}[V_\alpha] - 2K_3 = 0 \quad . \tag{2}$$

Similar equations can be obtained for the other five defect concentrations. Eq. 2 can be easily solved numerically to determine the defect concentration $[V_\alpha]$ for specified values of the equilibrium constants and composition ($x$). After one defect concentration is determined, corresponding values for the other five elementary defects are then most simply obtained using, once again, expressions for equilibrium constants from Table I. For example, $[B_\alpha]$ is given in terms of $[V_\alpha]$ by $[B_\alpha] = (K_{2A}/K_3)[V_\alpha]^2$.

*(e) Compounds with a unique or dominant type of defect disorder.* One type of defect combination often dominates. When the phase field extends to both sides of the stoichiometric composition, structural defects may be observable in annealed samples and used to identify low-energy elementary defects. Such structural defects are as a rule (but not always) constituents of the dominant thermally-activated defect combination. Exceptions may occur owing to differences in entropies of formation of different defect combinations. The entropy of formation $S_{vib}$, attributed to changes in vibrational degrees of freedom of the crystal that accompany formation of the defect combination, appears in the free energy of formation, $G = U + pV - TS_{vib}$, and leads to a prefactor in the equilibrium constant (e.g., $K = \exp(S_{vib}/k_B)\exp(-(U+pV)/k_BT)$). Presence of such prefactors modifies the behavior of solutions of concentration equations such as eq. 1, but are not explicitly considered here for simplicity. Exceptions to the rule can also occur when formation energies per defect of different defect combinations are accidentally close, as discussed below.

For intermetallic compounds having the CsCl structure, both antisite and triple-defect disorders have been observed.[23] For insulators, in which antisite atoms tend to be high-energy defects owing to coulomb interaction, Schottky and Frenkel disorders are more frequently observed[22,34,38] We now briefly examine eqs. 1 and 2 for situations in which antisite, Schottky, Frenkel or triple-defect types of disorder dominate.

(i) Antisite disorder. Assuming that only $A_\beta$ and $B_\alpha$ have significant concentrations, one can set $[V_\alpha] = [V_\beta] = [A_\tau] = [B_\tau] \cong 0$ in eq. 1. Then, using $[B_\alpha] = K_{2A}/[A_\beta]$ from Table I, eq. 1 reduces to a quadratic equation in $[A_\beta]$:

$$[A_\beta]^2 - 2x[A_\beta] - K_{2A} = 0. \tag{3}$$

(ii) <u>Schottky disorder</u>. Setting $[B_\alpha]=[A_\beta]=[A_\tau]=[B_\tau]\cong 0$ in eq. 1 or setting all $K$'s except $K_{2V}$ to zero in eq. 2 leads to a quadratic equation in $[V_\alpha]$:

$$(1-2x)[V_\alpha]^2 + 4x[V_\alpha] - (1+2x)K_{2V} = 0. \tag{4}$$

(iii) <u>Frenkel disorder</u>. Setting $[B_\alpha]=[A_\beta]=[V_\beta]=[B_\tau]\cong 0$ in eq. 1, or setting all K's except $K_{2FA}$ equal to zero in eq. 2, leads to a quadratic equation in $[V_\alpha]$ identical to eq. 4 except that the constant third term is replaced by $-3(1-2x)K_{2FA}$.

(iv) <u>Triple-defect disorder</u>. If the triple defect $2V_A+A_B$ is dominant, then setting all $K$'s to zero in eq. 2 except $K_3$ leads to the cubic equation

$$(1-2x)[V_\alpha]^3 + 4x[V_\alpha]^2 - 2K_3 = 0. \tag{5}$$

The corresponding equation for $[A_\beta]$ is also cubic:

$$([A_B]-2x)^2[A_B] - (\tfrac{1}{2}-x)^2 K_3 = 0. \tag{6}$$

Several different regimes of defect chemistry in ordered alloys are worth discussion.

1. <u>Zero temperature: structural defects</u>. As temperature decreases, all $K$'s go to zero, and, by definition, the equations yield concentrations of structural defects. For example, under either the Schottky, Frenkel or triple-defect models (eqs. 4 or 5), $[V_\alpha]=0$ for x>0 and $[V_\alpha]=-4x/(1-2x)$ for x<0. However, in the general case (eq. 2) it can be seen that the identity of structural defects will depend on ratios of equilibrium constants or, equivalently, on differences in formation energies. Thus it is even possible that a structural defect is not a constituent of the principal thermally-activated defect combination. To illustrate, consider eq. 2 while, for simplicity, neglecting equilibrium constants for Frenkel defects. It can be seen that if $G_3 < \tfrac{3}{2}G_{2A}$ and $G_3 < \tfrac{3}{2}G_{2V}$, the triple defect will be thermally activated at the stoichiometric composition. If, in addition, however, $G_3 > G_{2A}$, then the first term in eq. 2 diverges at low temperature and, as a consequence, the antisite pair will be the structural defect. Specifically, this means that the dominant elementary defect for A-poor compositions at low temperature will be $B_\alpha$ but that, with increasing temperature, $[B_\alpha]$ will decrease and $[V_\alpha]$ will increase at the same composition. This scenario was described recently by Ren and Otsuka in their analysis of the x-ray and neutron measurements on NiAl and similar systems by Kogachi and coworkers, summarized in ref. 39. However, when $G_3 < G_{2A}$ and $G_3 < 2G_{2V}$, the triple defect will be thermally activated and its constituents ($A_\beta$ and $V_\alpha$) will also be structural defects on either side of the stoichiometric composition. This last scenario was not discussed in the recent work of those researchers.[31,39]

2. <u>Stoichiometric regime</u>. The stoichiometric regime is defined as the composition range for which |x| is much less than any relevant defect concentration. Eqs. 3-6 then all reduce to

expressions for defect concentrations that have an Arrhenius temperature dependence. For example from eq. 4, $[V_\alpha] = K_{2V}^{1/2} = \exp(-G_{2V}/2k_BT)$ at x=0. Since defect concentrations increase with temperature, the range of the stoichiometric regime likewise increases with temperature. When equilibrium defect concentrations are small, either due to high formation energies or low temperature, it may be practically difficult to make measurements strictly in the stoichiometric regime due to uncertainties in the mean composition or to sample inhomogeneity.

3. <u>Far from stoichiometry</u>. When |x| is much greater than the relevant defect concentration, then higher-order terms in defect concentrations in equations such as eqs. 2-6 can be dropped, leading to qualitatively different behavior than at stoichiometry. For example, from eq. 4 for the vacancy pair, $[V_\alpha] \cong K_{2V}/4x = \exp(-G_{2V}/k_BT)/4x$ at large, positive x. For negative x, on the other hand, a large concentration of order 4|x| of structural $V_\alpha$ remains.

Tab.II

Expressions for selected defect concentrations at stoichiometry (x=0) and far from stoichiometry were derived using eqs. 1-6 and similar equations for other types of disorder. These are shown in Table II for the five types of defect disorder listed of Table I. As can be seen, all defect concentrations at the stoichiometric composition are thermally activated with an effective activation energy given by the free energy of the defect combination divided by the number of constituent defects, e.g., the effective activation energy for $A_\beta$ is $G_{2A}/2$ or $G_3/3$, depending on the defect combination that is assumed to be dominant. As can also be seen, the effective activation energy differs for compositions far from stoichiometry; e.g., the effective activation energy for $A_\beta$ becomes $G_{2A}$ or $G_3$ for A-poor compositions. Such differences in effective activation energy have been observed in PAC experiments[40] and elsewhere. For a general nonstoichiometric composition, the temperature dependence of concentrations of elementary defects of course does not a simple thermally activated behavior. However, one can use measured concentrations and the known composition (x) in conjunction with expressions for a particular defect model such as eqs. 3-6 to determine the appropriate *K*. This approach was used in ref. 40 to determine the formation energy of the triple-defect combination in NiAl.

*(f) Defect energies*. While the formation energy of a defect combination can be determined from an Arrhenius plot of its equilibrium constant, the formation energies of its elementary point defects by themselves cannot be measured. Individual defect energies are thus curiously 'hidden' but remain coupled through the energies of defect combinations to which they contribute. Defining energies of defect combinations as sums of energies of elementary defects is necessary to ensure that energies of the defect combinations are consistently defined (see eqs. 7-11 below.).

The only way to determine individual defect energies appears to be through computation. They are defined here with respect to the energy of the same site in the defect-free lattice, independent of any deviation from stoichiometry. Thus, the energy of an antisite $A_\beta$ atom, written $G(A_\beta)$, equals the difference between energies of a crystal containing an $A_\beta$ defect and a $B_\beta$ atom. The B-atom can be thought of as being removed to infinity, from where the A-atom is brought in. Similarly, the energy of vacancy $V_\alpha$ is defined as the difference between the energy of the crystal with the vacancy and with a host $A_\alpha$ atom, and is normally quite positive owing to breaking of bonds around the removed A-atom. Finally, the energy of an interstitial defect $A_\tau$ is obtained by bringing an A-atom in from infinity and placing it on an empty interstitial site, and may be quite negative owing to establishment of bonds around the introduced atom. With this convention, site energies of substitutional host atoms $A_\alpha$ and $B_\beta$ are zero by definition, $G(A_\alpha) = G(B_\beta) = 0$. Solute energies on different sites are similarly defined. Formation energies of defect combinations are equal to the sums of energies of the constituent defects, with the energy of formation of new unit cells included, if any. Formation energies of the five fundamental defect combinations in an AB compound are defined in terms of individual defect energies as follows:

$$G_{2V} = G(V_\alpha) + G(V_\beta) + G_{cell}, \qquad (7)$$

$$G_{2A} = G(A_\beta) + G(B_\alpha), \qquad (8)$$

$$G_3 = 2G(V_\alpha) + G(A_\beta) + G_{cell}, \qquad (9)$$

$$G_{2FA} = G(V_\alpha) + G(A_\tau), \qquad (10)$$

$$G_{2FB} = G(V_\beta) + G(B_\tau). \qquad (11)$$

The energy to form a unit cell, $G_{cell}$, is equal to the (negative) cohesive energy per unit cell of the perfect compound. An individual defect energy may be negative as long as free energies of all defect combinations to which it belongs are positive. A possible example of a defect with an energy close to zero or negative is the antisite defect $Ni_{Al}$ in NiAl. This is because the Ni-antisite defect fits readily in the volume of the Al-atom it replaces and because the bonding of the Ni-antisite atom with its eight Ni neighbors is comparable to the bonding of an Al atom in the same location.

A discussion of particular methods for calculating defect energies is beyond the scope of this paper. The formalism described above is general, however, and almost any method could be used to calculate the energies. A simple method for calculating defect energies is the near-neighbor bond model which has been used extensively for intermetallics.[23,26,27,31] In the perfect CsCl structure, atoms of one type are surrounded by eight atoms of the other type. There is a cohesive energy of four AB bonds per atom, each bond having an associated (negative) energy $U_{AB}$, leading to a cohesive energy $G_{cell} = 8U_{AB}$ for the two atoms in the unit cell. Considering only vacancy and antisite defects, one can assign energies $U_{AA}$, $U_{BB}$, $U_{VA}$ and $U_{VB}$ to bonds between near neighbor host atoms and to 'ghost bonds' between host atoms and neighboring vacancies, making for a total of five bond energies. The four site energies $G(V_\alpha)$, $G(V_\beta)$, $G(A_\beta)$, and $G(B_\alpha)$ and cohesive energy $G_{cell}$ of the present site-energy can be expressed equivalently in terms of those five bond energies, so that the site-energy and bond models can be seen to be formally equivalent.

Different types of disorder will coexist if free energies per defect of different combinations are nearly equal. In that case two or more equilibrium constants in an equation such as eq. 2 will have significant non-zero values. Numerical analysis of such systems using eq. 2 or corresponding expressions for other defects and structures is straightforward. As a rule of thumb, simulations have typically shown that energy differences per defect of order 0.1 eV are usually sufficient to make the combination dominant that has the lowest energy per defect. Since formation energies are typically greater than 1 eV, the general likelihood that different types of disorder coexist is small.

### III. SITE PREFERENCES OF SOLUTES IN AB COMPOUNDS

Consider solute S dissolved in a binary compound of composition $A_{1+2x}B_{1-2x}$. The solute will sit on sublattices $\alpha$, $\beta$, and $\tau$ with an equilibrium distribution established by reactions that transfer S between sites. Between substitutional sites $\alpha$ and $\beta$ one can envision three distinct transfer reactions, with equilibrium constants expressed in terms of the fractional concentrations of reactant and product species as follows:

$$S_\beta + A_\alpha \rightarrow S_\alpha + A_\beta, \qquad K_a = \frac{[S_\alpha]}{[S_\beta]}[A_\beta] = \exp(-G_a/k_BT), \qquad (12)$$

$$S_\beta + V_\alpha \rightarrow S_\alpha + V_\beta, \qquad K_b = \frac{[S_\alpha][V_\beta]}{[S_\beta][V_\alpha]} = \exp(-G_b/k_BT), \qquad (13)$$

$$S_\beta + B_\alpha \rightarrow S_\alpha + B_\beta, \qquad K_c = \frac{[S_\alpha]}{[S_\beta]}[B_\alpha]^{-1} = \exp(-G_c/k_BT). \qquad (14)$$

Transfer reactions between substitutional and interstitial sites include the following two, among others:

$$S_\alpha \rightarrow V_\alpha + S_\tau, \qquad K_d = \frac{[S_\tau]}{[S_\alpha]}[V_\alpha] = \exp(-G_d/k_BT), \qquad (15)$$

$$S_\beta \rightarrow V_\beta + S_\tau, \qquad K_e = \frac{[S_\tau]}{[S_\beta]}[V_\beta] = \exp(-G_e/k_BT). \qquad (16)$$

It will be assumed that the solute concentration is sufficiently dilute that intrinsic defect concentrations are undisturbed by the site preference of the solute. Then, intrinsic defect concentrations in the above five equations can continue to be derived by solution of eq. 2 or corresponding expressions for other defects. Outlined in an appendix is the extension of the model for non-negligible concentrations of solute, which is straightforward but laborious.

The measure of preference of a solute for one site over another is taken to be the ratio of site fractions for the two sites. Signal amplitudes measured using hyperfine interaction methods are directly proportional to site fractions, or defect concentrations. For a general crystal structure, the site fraction of solute S on sublattice $\alpha$ of a compound is $f_\alpha = N_\alpha[S_\alpha]/N_{tot}^S$, in which $N_\alpha$ is the number of $\alpha$-sites, and $N_{tot}^S$ is the total number of solutes in the compound. The ratio of site fractions on two general sublattices $\beta$ and $\alpha$ is then

$$R_\alpha^\beta \equiv \frac{f_\beta}{f_\alpha} = \frac{N_\beta [S_\beta]/N_{tot}^S}{N_\alpha [S_\alpha]/N_{tot}^S}. \tag{17}$$

If two sublattices have the same number of sites ($N_\alpha = N_\beta$), the site-fraction ratio reduces trivially to the ratio of solute concentrations:

$$R_\alpha^\beta = \frac{f_\beta}{f_\alpha} = \frac{[S_\beta]}{[S_\alpha]}. \tag{18}$$

Site-fractions can be calculated directly from the site-fraction ratios. For example,

$$f_\beta = \frac{1}{1 + 1/R_\alpha^\beta + 1/R_\tau^\beta}. \tag{19}$$

Tab. III

In Table III are listed selected expressions for site-fraction ratios obtained using eq. 17 with eqs. 12-16. Expressions for activation energies of solute transfer reactions in terms of defect and solute energies are also given in the table. As a check, the energy expressions as defined lead to appropriate limiting cases when $S=A$ or $S=B$.

The ratio of fractional concentrations of solute S on the two sites, $[S_\alpha]/[S_\beta]$, appears in each of eqs. 12-14, so that $R_\alpha^\beta$ can be expressed in three alternative ways:

$$R_\alpha^\beta = \frac{[S_\beta]}{[S_\alpha]} = [A_\beta] \exp(+G_a/k_B T) = \frac{[V_\beta]}{[V_\alpha]} \exp(+G_b/k_B T) = [B_\alpha]^{-1} \exp(+G_c/k_B T). \tag{20}$$

Eq. 20 and entries in Table III are quite interesting:

1. Each of the three expressions in eq. 20 is proportional to an intrinsic defect concentration or ratio of concentrations: $[A_\beta]$, $[V_\beta]/[V_\alpha]$ or $[B_\alpha]^{-1}$. Indeed, the entire composition dependence is contained in the concentrations. The site-fraction ratio depends strongly on $x$ through this dependence and, like the concentrations, will in general not have a simple thermally-activated form. Each expression in eq. 20 also contains a Boltzmann factor with an energy equal to the sum of a solute site-energy difference and the energy of the corresponding intrinsic defect (cf. Table III.) Identical expressions for $R_\alpha^\beta$ in terms of $[A_\beta]$ or $[B_\alpha]^{-1}$ and with Boltzmann factors whose energies are given in lines 1 and 2 of Table III were obtained earlier by Woodward et al. (ref. 18, equations 10 and 11.) The proportionality of $R_\alpha^\beta$ with $[V_\alpha]/[V_\beta]$ and expressions given above for site-fraction ratios of substitutional and interstitial solutes have not to our knowledge been reported before.

2. As the composition changes from x<0 to x>0, the dominant elementary defects change from structural defects of one type ($V_\alpha$, $B_\alpha$, $B_\tau$) to the other ($V_\beta$, $A_\beta$, $A_\tau$). Consideration of the positions of defect concentrations in numerators or denominators of expressions for site-fraction ratios shows that β-sites are favored by solutes in B-poor alloys where the concentration of $V_\beta$, $A_\beta$, or $A_\tau$ is large, and α-sites by solutes in A-poor alloys where the concentration of $V_\alpha$, $B_\alpha$, or $B_\tau$ is large.

3. The site fractions of solutes on substitutional sites vary monotonically with composition. For example, from eq. 19 and Table III, $f_\beta \sim 1/(1 + c_1/[A_\beta] + c_2/[V_\beta])$. Note that the two defects are alternative structural defects and that the concentrations of such defects are monotonic in each other; for example, from Table I, $[A_\beta]^2 = (K_3/K_{2V}^2)[V_\beta]$. Since concentrations of all intrinsic defects vary monotonically with x, the site-fraction must similarly be monotonic in x. The site fraction for solutes on interstitial sites, $f_\tau = 1/(1 + R_\tau^\alpha + R_\tau^\beta) \sim 1/(1 + c_3[V_\alpha] + c_4[V_\beta])$ will be maximum near the stoichiometric composition, where concentrations of the two vacancies (opposing structural defects) are least.

4. Since alternative expressions for $R_\alpha^\beta$ in eq. 20 are equal, finite concentrations of all four intrinsic defect species appearing in the equation must be present in thermal equilibrium, no matter how small. This emphasizes the existence of minority defect species in real systems. Equality between the alternative expressions implies a high degree of coupling among defect and solute concentrations that probably still rmains valid for conditions out of equilibrium, such as in quenched samples. From a practical point of view, one can calculate $R_\alpha^\beta$ using whichever expression is most convenient in terms of knowledge of defect concentrations and energies.

5. Finite experimental precision limits the ability to determine very large and small site-fraction ratios. For example, an experimental uncertainty in site fractions of 1% limits the dynamic range of site-fraction ratios to between 0.01 and 100. Actual ratios may easily fall far outside this range, especially in samples equilibrated at low temperature.

6. Activation energies $G_{a-e}$ for solute transfer reactions have a profound effect on the scale of site-fraction ratios. In particular, the signs of $G_{a-e}$ determines whether a ratio is much greater or less than one at low temperature and therefore helps to determine the solute site preference.

Trends in the dependences of site-fraction ratios on *x* can be worked out under different assumptions about which defect combination is dominant. Using expressions for defect concentrations from Table II and the solute reactions in eq. 12-16, one can derive expressions for site-fraction ratios for x<<0, x=0 and x>>0. Selected expressions are given in Table IV and depend on equilibrium constants for defect formation and solute transfer. Depending on values of the *K's*, the site-fraction ratios may be much greater or less than one. Examination of entries such as in Table IV allows one to specify conditions under which particular site fractions will dominate for all compositions or whether the principal site preference of a solute will change from one side of the stoichiometric composition to the other. To illustrate, Table V gives conditions under which solutes will always reside on the α-sublattice or β-sublattice or will switch sites (under the assumption that concentrations of host and solute interstitials can be

[Tab. IV]

[Tab. V]

neglected.) Also given is the condition under which site fractions are equal for the stoichiometric composition. The entries were obtained by examining entries in the three corresponding lines of Table IV in the low-temperature limit. For example, assume the triple-defect is dominant. The solute will always sit on the α-sublattice when the difference $\Delta G \equiv G(S_\alpha) - G(S_\beta)$ of energies of a solute on the two sites is such that $\Delta G < -G(A_B)$. Likewise, the solute will always sit on the β-sublattice if $G(S_A)$ is sufficiently greater than $G(S_B)$ that $\Delta G > 2G(V_A)$. However, the solute will switch from the β-sublattice in A-rich compounds to the α-sublattice in B-rich compounds if $-G(A_B) < \Delta G < 2G(V_A)$.

Fig. 2

The dependence of $R_\alpha^\beta$ on composition and temperature is simulated for the triple-defect model in Figures 2 and 3. For these simulations, the solute is assumed to only populate substitutional sites. The calculated trends will be fairly accurate for compositions close to stoichiometry, say within 1 at.%, but are included out to x=-0.1 and +0.1 to show qualitative trends and the influence of temperature. For both figures, the value $G_3$= 1.6 eV was assumed, consistent with results of measurement for NiAl,[23,40] and formation energies of all other defect combinations were assumed to be much higher. [$A_\beta$] was calculated using eq. 6 and then $R_\alpha^\beta$ was calculated from its expression in terms of [$A_\beta$] in eq. 20. Figure 2 shows a log-plot of the site-fraction ratio versus composition at T= 600 K for various values of the transfer reaction energy $G_a$ between 0 and 2.0 eV. The value $R_\alpha^\beta = 1$ is indicated by the dashed line. Examination of Fig. 2 leads to the following observations:

1. A step-like discontinuity in $R_\alpha^\beta$ of magnitude $\sim 10^9$ is observed as one crosses the stoichiometric composition from x= -0.01 (49 at.% A) to +0.01 (51 at.% A). The magnitude of the step can be interpreted using Table IV by dividing the site-fraction ratio entry for x>>0 by the entry for x<<0, which for the triple defect gives $32|x^3|K_3^{-1}$. Similar results can be obtained from Table IV for antisite or Schottky defects. Thus, the magnitude of the step gives a measure of the equilibrium constant for the dominant defect combination.

2. $G_a$ is equal to the energy of the antisite defect $A_\beta$ plus the difference in site-energies of the solute atom. For $G_a > G_3$(=1.6eV), $R_\alpha^\beta$ is always much greater than one, so that $f_\beta \sim 1$. For $G_a < 0$ eV, $R_\alpha^\beta$ is always much smaller than one, so that $f_\alpha \sim 1$. For intermediate values, $0 < G_a < G_3$, the dominant site-fraction crosses from $f_A$ for x<0 to $f_B$ for x>0. Thus, whether or not there is a change in site preference as a function of the composition depends on the relative magnitude of $G_a$ and $G_3$, as listed in the row for the triple defect in Table V. .

3. The composition dependence of $R_\alpha^\beta$ shown in Fig. 2 depends only on the concentration of intrinsic defects and a thermally activated scaling factor. As observed, $R_\alpha^\beta$ is greater in A-rich than in A-poor alloys, as noted earlier. This behavior again supports a rule of thumb that solutes have a propensity to occupy sites of the more-deficient element.

Fig. 3

Figure 3 shows how site preferences change with temperature for fixed energy parameters $G_3$= 1.6 eV and $G_a$= 1.0 eV. In the figure, the log of $R_\alpha^\beta$ is plotted versus composition for five temperatures between 300 and 1500 K. Examination of Fig. 3 leads to several observations.

1. The magnitude of the step-like discontinuity in $R_\alpha^\beta$ decreases with increasing temperature. With increasing temperature, sites become more equally occupied, so that $R_\alpha^\beta$ is observed to decrease in A-rich alloys (x>0) and increase in A-poor alloys (x<0).

2. With increasing temperature, the step-like discontinuity becomes broader as a consequence of the increasing concentration of thermal defects near stoichiometry.

3. Site preferences can change with temperature. At the composition x~ -0.01, for example, the site preference is observed to change from α-site at low temperature to β-site at high temperature.

Fig. 4

Figure 4 shows site fractions $f_\alpha$ and $f_\beta$ calculated from the graph of the site-fraction ratio for 1200 K in Fig. 3. A number of qualitative features are noted. As can be seen, the site preference changes from site α for x<-0.015 to site β for greater values of x. Substitutional site fractions are found to always increase or decrease monotonically as a function of composition. Site fractions may be close to 0 and 1, as shown for x>0 in the figure, or have intermediate values, as shown for x<0. Finally, it should be noted that the composition at which a site fraction crosses over will not occur precisely at the stoichiometric composition in samples equilibrated at elevated temperature.

Fig. 5

Figure 5 shows a different simulation in which the triple defect is dominant but in which the energies are such that the solute appreciably populates sites α, β and τ. The site fraction of solutes on the interstitial site is observed to have a sharp peak very near the stoichiometric composition. As explained above, this is because the site fraction $f_\tau \equiv 1/(1 + R_\tau^\alpha + R_\tau^\beta)$ is maximum when the sum of the two site-fraction ratios appearing in the equation (proportional to $[V_\alpha]$ and $[V_\beta]$) are minimum. The sum is minimum near stoichiometry. The maximum value of the interstitial site fraction may occur at a composition displaced somewhat from stoichiometry due to different magnitudes of the two site-fraction ratios (cf., e.g., Fig. 8 below.) The substitutional site fractions continue to vary monotonically with composition.

## IV. DEFECTS IN $Ni_2Al_3$ COMPOUNDS

The $Ni_2Al_3$ structure is shown schematically in Fig. 1. It is closely related to the CsCl structure[41] and can be thought of as arising from Ni-poor NiAl by condensation of structural Ni-vacancies on every third 111 Ni-plane.[42] The ordering of vacancies is accompanied by a slight contraction (~1%) of the crystal along the normal 111 axis and slight movements of 111 lattice planes of Ni and Al atoms toward the empty Ni-planes. The contraction and movements are all small so that the crystal structure can still be referenced to the CsCl-structure from which it is derived. Most compounds having the $Ni_2Al_3$ structure are of type $A_2B_3$, with A= (Ni, Pd, Pt) and B= (Al, Ga, In).

Apart from its intrinsic interest, analysis of defects and site preferences in $A_2B_3$ is a step towards consideration of more complex structures. Of six elementary 111-planar sublattices, one is the empty A-sublattice, with sites designated X below, two are occupied by A-atoms and are equivalent (site $\alpha$), two other equivalent sublattices are occupied by B atoms and adjacent to the empty planes (site $\beta2$), and one sublattice is occupied by B-atoms and is more distant from the empty planes (site $\beta1$). Thus, there are four distinct sites for defects or solutes. As will become clear, there are many similarities as well as important differences in site-preference behavior of the two structures.

Elementary intrinsic defects will be assumed to be vacancies and antisite atoms on the three substitutional lattice sites $\alpha$, $\beta1$ and $\beta2$ and "interstitials" $A_x$ and $B_x$ on sublattice X, for a total of 8 elementary defects. The relative numbers of sites are $\alpha(2)$, $X(1)$, $\beta1(1)$, $\beta2(2)$ and the defects are $B_\alpha$, $V_\alpha$, $B_x$, $A_{\beta1}$, $A_{\beta2}$, $V_{\beta1}$, $V_{\beta2}$, and $A_x$. Fractional defect concentrations are defined with respect to the total numbers of sites on the sublattice(s), which differ by factors of 1 or 2

Equation of constraint among defect concentrations in $A_{2+5x}B_{3-5x}$. Deviations from the stoichiometric composition of $A_{2+5x}B_{3-5x}$, 40 at.% A, are again measured by x, with, for example, x= +0.01 for 41 at.% A and x=-0.01 for 39 at.% A. The equation of constraint among defect concentrations can be shown by methods of the Appendix to be

$$2[B_\alpha] + (\tfrac{6}{5} - 2x)[V_\alpha] + (\tfrac{2}{5} + x)[B_x] + 5x = [A_{\beta1}] + 2[A_{\beta2}] + (\tfrac{2}{5} + x)([V_{\beta1}] + 2[V_{\beta2}]) + (\tfrac{3}{5} - x)[A_x]. \quad (21)$$

In eq. 21, alternative structural defects in off-stoichiometric alloys are again collected on the same sides of the equation. These are, respectively, $B_\alpha$, $V_\alpha$ and $B_x$ for A-poor and $A_{\beta1}$, $A_{\beta2}$, $V_{\beta1}$, $V_{\beta2}$ and $A_x$ for A-rich alloys. Solute concentrations are assumed to be negligible in comparison and are neglected in eq. 21.

A simplification occurs if energies of defects (but not solutes) on the $\beta1$ and $\beta2$ sublattices are assumed to be equal, so that $[V_{\beta1}] \equiv [V_{\beta2}]$ and $[A_{\beta1}] \equiv [A_{\beta2}]$. One can then treat the $\beta1$ and $\beta2$ sublattices to be equivalent insofar as intrinsic defects are concerned. Average concentrations of defects over both sublattices are then given by:

$$[V_\beta] \equiv \frac{[V_{\beta1}]}{3} + \frac{2[V_{\beta2}]}{3}, \qquad [A_\beta] \equiv \frac{[A_{\beta1}]}{3} + \frac{2[A_{\beta2}]}{3}. \qquad (22)$$

In this situation, the equation of constraint in eq. 21 reduces to

$$2[B_\alpha] + (\tfrac{6}{5} - 2x)[V_\alpha] + (\tfrac{2}{5} + x)[B_X] + 5x = 3[A_\beta] + (\tfrac{6}{5} + 3x)[V_\beta] + (\tfrac{3}{5} - x)[A_X]. \qquad (23)$$

<u>Defect combinations and formation reactions</u>.  As for CsCl, defects only form in combinations that preserve the composition of the alloy.  With 8 elementary defects, of which 3 and 5 are structural defects on the two sides of stoichiometry, there is a total of 15 defect combinations that involve two elementary defects.  If defects on the two β sublattices, $V_{\beta 1}$ and $V_{\beta 2}$, and $A_{\beta 1}$ and $A_{\beta 2}$, are assumed to have the same energies, the number of distinct elementary defects is reduced to 6 and the number of distinct defect combinations to 9.  In this approximation, formation reactions for all nine defect combinations are listed in Table VI.  The first five defect combinations lead to coupling among all defect concentrations and are adopted as more fundamental.  These include a 5-vacancy Schottky defect, formed heuristically by removing a molecule from the interior and placing it on the surface, creating $2V_\alpha$ and $3V_\beta$ vacancies and increasing the number of unit cells by one.  The second combination is the antisite atom pair.  Third is the 8-defect, a defect comprised of $V_\alpha$ and $A_\beta$ defects that is the analog of the triple defect in the CsCl structure. The fourth and fifth defect combinations are Frenkel defect pairs formed by displacing an A or B atom from the α or β sublattice to the X sublattice.  For a general stoichiometry $A_aB_b$ with only one distinct site for each element, the elementary defects can be taken to consist of a generalized Schottky defect with (a+b) vacancies, the antisite defect, the two Frenkel defects, and one of two mixed vacancy-antisite analogs of the triple-defect in the CsCl structure, containing a total of (a+b) vacancies and either a or b antisite atoms.

The second group of four defect combinations in Table VI can be derived from the first five.  For example, the interstitial atom pair can be created by formation of two Frenkel-A and three Frenkel-B defects, followed by annihilation of a 5-vacancy Schottky defect.  Three 7-defects can be created by forming five 5-vacancy defects and three antisite pairs, followed by annihilation of two 8-defects.  The anti-8 defect can be formed from five Frenkel-A defects and three antisite pairs by annihilation of an 8-defect. Antidefect pairs for $Ni_2Al_3$ or any other binary phase are the same as those for the CsCl structure (with X taking the role of τ): $(A_\beta,B_\alpha)$, $(A_x,V_\alpha)$, $(V_\beta,B_x)$.  Note that the anti-8-defect can also be obtained from the formation reaction of an 8-defect by transforming each defect into its antidefect, whence its name.

Dominant defect combinations could conceivably include more than two types of elementary defects.  At the bottom of Table VII are listed two examples of combinations involving three defects: a 3-defect and its antidefect.  Such defects can also be formed out of the five fundamental reactions.  For example, nine anti-3-defects can be formed by nine Frenkel-B plus three 8-defect reactions, less three 5-vacancy reactions, with a formation energy per 3-defect equal to $(9G_{2FB} + 3G_8 - 3G_{5V})/9$. Such higher-order combinations really amount to coexistence of two or more binary defect combinations and therefore need not be considered apart.

<u>Additional reactions for distinct β1 and β2 sublattices</u>.  When energies of antisite defects and vacancies differ between the two β-sublattices, additional formation reactions are necessary to couple the defect concentrations.  For the $Ni_2Al_3$ structure, the fundamental reactions involving defects on the β-sublattice have to be doubled to include the new β2 sublattice, leading to a total of nine fundamental reactions in the more general structure.

<u>Solving for defect concentrations in the general case</u>.  The five fundamental formation reactions in Table VI and four additional ones for the second β-sublattice can be used to eliminate all defect concentrations but one from the equation of constraint (eq. 21). The explicit result for concentration $[V_\alpha]$ is as follows:

$$2\frac{K_{2A1}}{K_{81}^{1/3}}[V_\alpha]^{10/3} + (\tfrac{6}{5} - 2x)[V_\alpha]^{8/3} + (\tfrac{2}{5} + x)\frac{K_{5I}^{1/3}}{K_{2FA}^{2/3}}[V_\alpha]^{7/3} + 5x[V_\alpha]^{5/3}$$
$$- (\tfrac{2}{5} + x)(K_{5V}^{1/3} + 2K_{5V2}^{1/3})[V_\alpha] - (\tfrac{3}{5} - x)K_{2FA}[V_\alpha]^{2/3} - K_{81}^{1/3} - 2K_{82}^{1/3} = 0 \qquad (24)$$

in which equilibrium constants for the set of four formation reactions involving defects on the β2-sublattice have trailing "2"-subscripts. Eq. 24 can be solved for $[V_\alpha]$ for given values of the eight $K's$ and x, after which the other eight defect concentrations can be obtained using the nine formation reactions.

The binary defect combination having the smallest formation energy per defect will normally be comprised of structural defects on the two sides of stoichiometry and will be the dominant thermal defect. A more detailed analysis of eq. 24 and of analogous equations for the other defect concentrations can be made to obtain more precise criteria for dominance of a defect combination.

<u>The dominant defect combination in Ni$_2$Al$_3$.</u> Most CsCl phases formed from A=(Ni, Pd, Pt) and B=(Al, Ga, In) have as structural defects A-vacancies ($V_\alpha$) and A-antisite atoms ($A_\beta$), and the equilibrium defect is the triple-defect. Since the local surroundings of atoms do not differ much between CsCl and Ni$_2$Al$_3$ phases, this suggests that the dominant defects in Ni$_2$Al$_3$ phases might also be composed of $A_\beta$ and $V_\alpha$ defects, that is, the 8-defect. Instead, experimental evidence supports the anti-8-defect model for Ni$_2$Al$_3$ rom combined measurements of lattice parameters and mass densities by Taylor and Doyle carried out in the same way that NiAl was studied by Bradley and Taylor some 35 years earlier. Ni$_2$Al$_3$ has a phase range extending from about 38 to 42 at.% Ni, so that structural defects in both Ni-rich and Ni-poor samples are accessible to measurement. Taylor and Doyle found best agreement for structural defects that were Ni-atoms in empty Ni-planes in Ni-rich alloys ($A_X$) and Al-antisite atoms in Ni-poor alloys ($B_\alpha$). This suggests that the anti-8-defect has a lower formation energy than the 8-defect. However, this conclusion may have been premature. In order to reduce porosity prior to making density measurements, Taylor and Doyle compacted their samples under a pressure of 56 kbar. Such a high pressure could have converted equilibrium defects into metastable types. In particular, rearrangement of the formation reaction for the anti-8-defect in Table VI in two different ways leads to the transformation reactions $5V_\alpha \rightarrow 3B_\alpha - uc$ and $3A_\beta \rightarrow 5A_X - uc$. Each reaction would be driven to the right-hand side by application of pressure through the loss of a unit-cell volume, equal to about 0.25 nm$^3$. Thus, an applied pressure of 56 kbar would result in an enthalpy difference of order $p\Delta V \approx -8\,eV$ per reaction, or about 2 eV per converted defect. Such an appreciable enthalpy might have converted $V_\alpha$ and $A_\beta$ defects, respectively, into metastable $B_\alpha$ and $A_X$. Such conversions were studied in other systems by Taylor and Doyle.[43]

<u>Other evidence for identifying structural defects in Ni$_2$Al$_3$.</u> Hyperfine interaction measurements provide additional information about structural defects. PAC measurements on indium probes in Ni$_2$Al$_3$ have been carried out at ambient pressure for compositions over the range 38-42 at.% Ni. For Ni-poor compositions, PAC signals of probe atoms become highly broadened, indicating that the indium probes are eliminated from all regular lattice sites, so that no information is available for x<0. For stoichiometric and Ni-rich samples, the dominant signal is a unique quadrupole interaction that can be attributed unambiguously to indium solutes on β2 sites.[20] For compositions near 42 at.% Ni, the β2 PAC signal is broadened owing to structural defects. The actual broadening observed for Ni-rich samples is quite small, indicating that structural defects are not located on sites next to the probes. Of possible defects in A-rich alloys ($A_{β1}$, $A_{β2}$, $V_{β1}$, $V_{β2}$, $A_x$), the small broadening appears to rule out $A_X$, which would be expected to be located on near-neighbor sites to the $In_{β2}$ probes and to produce a strong disturbance. (This assumes that $A_X$ defects are not 'repelled' from positions next ot the probe.) On the other hand, possible antisite $A_β$ atoms would be located in second-neighbor positions and produce only weak quadrupole interactions, as observed. Thus, the PAC measurements do not support Taylor and Doyle's evidence for the anti-8-defect model and are consistent with the idea that the 8-defect is the dominant defect.

Simulation of defect concentrations. Defect concentrations in $A_{2+5x}B_{3-5x}$ were calculated as a function of *x* for T= 1500 K using a set of formation energies arbitrarily chosen that made the 8-defect dominant. Specific values used for defect energies were $G(V_\alpha)$, $G(V_\beta)$, $G(A_\beta)$, $G(B_\alpha)$, $G(A_\tau)$, $G(B_\tau)$, and $G_{cell}$ were, respectively, 4, 8, 2, 4, -1, -1, and -22.14 eV, from which the energies of defect combinations $G_{5V}$, $G_{2A}$, $G_8$, $G_{2FA}$, $G_{2FB}$, $G_{5I}$, $G_7$, $G_{8X}$ and $G_{7X}$ are, respectively, 10.1, 6.0, 4.4, 3.1, 7.0, 17.1, 25.9, 29.1 and 21.1 eV. The defect combination having the least formation energy per constituent elementary defect is the 8-defect, at 0.55 eV per defect. First, eq. 24 was solved for $[V_\alpha]$, after which the other concentrations were obtained using mass-action relations. Concentrations obtained are plotted versus x in a linear plot (Figure 6) and log-plot (Figure 7). Principal defects observed are $V_\alpha$ and $A_\beta$, constituents of the 8-defect. These are the structural defects that would remain at low temperature in A-poor and A-rich alloys, respectively. Also visible in the plots is a thermally-activated concentration of $B_X$ and lesser activated concentrations of $B_\alpha$, $V_\beta$ and $A_X$. At stoichiometry, one observes thermally activated concentrations of $V_\alpha$, $B_X$ and $A_\beta$ equal to 1.67, 0.50 and 1.53 %, respectively. In Fig. 7 it can be seen that alternative structural defects in A-poor ($V_\alpha$, $B_X$, $B_\alpha$) and A-rich alloys ($A_\beta$, $V_\beta$, $A_X$) have similar monotonic S-shape composition dependences. More quantitatively, concentration curves for defects and antidefects are observed to be reciprocal apart from scaling factors. Compared with Fig 5, for example, changes of concentration with composition near x=0 are fairly gradual due to the high simulation temperature used.

## V. SITE PREFERENCE OF SOLUTES IN $Ni_2Al_3$ COMPOUNDS

With four solute sites (α, X, β1 and β2) there are six distinct site-fraction ratios and many possible transfer reactions. Consider first reactions that establish equilibrium between solutes on the β1 and α sublattices. In analogy with eqs. 12-14, there are three reactions involving vacancies or antisite atoms, with equilibrium constants as follows:

$$S_{\beta 1} + A_{\alpha} \rightarrow S_{\alpha} + A_{\beta 1}, \qquad K_{a1} = \frac{[S_{\alpha}]}{[S_{\beta 1}]}[A_{\beta 1}] = \exp(-G_{a1}/k_B T); \qquad (25)$$

$$S_{\beta 1} + V_{\alpha} \rightarrow S_{\alpha} + V_{\beta 1}, \qquad K_{b1} = \frac{[S_{\alpha}]}{[S_{\beta 1}]}\frac{[V_{\beta 1}]}{[V_{\alpha}]} = \exp(-G_{b1}/k_B T); \qquad (26)$$

$$S_{\beta 1} + B_{\alpha} \rightarrow S_{\alpha} + B_{\beta 1}, \qquad K_{c1} = \frac{[S_{\alpha}]}{[S_{\beta 1}]}[B_{\alpha}]^{-1} = \exp(-G_{c1}/k_B T). \qquad (27)$$

One finds from eqs. 25-27 that the ratio of site fractions of solutes on the β1- and α–sublattices is given by

$$R_{\alpha}^{\beta 1} \equiv \frac{f_{\beta 1}}{f_{\alpha}} = \frac{[S_{\beta 1}]}{2[S_{\alpha}]} = \tfrac{1}{2}[A_{\beta 1}]\exp(+G_{a1}/k_B T) = \frac{[V_{\beta 1}]}{2[V_{\alpha}]}\exp(+G_{b1}/k_B T) = \tfrac{1}{2}[B_{\alpha}]^{-1}\exp(+G_{c1}/k_B T),$$
(28)

quite similar to eq. 20 for the CsCl structure apart from factors of 2 that account for the 2:1 ratio of sites on the α- and β1-sublattices. For equilibration of solutes on β2 and α sublattices, one similarly obtains:

$$R_{\alpha}^{\beta 2} \equiv \frac{f_{\beta 2}}{f_{\alpha}} = \frac{[S_{\beta 2}]}{[S_{\alpha}]} = [A_{\beta 2}]\exp(+G_{a2}/k_B T) = \frac{[V_{\beta 2}]}{[V_{\alpha}]}\exp(+G_{b2}/k_B T) = [B_{\alpha}]^{-1}\exp(+G_{c2}/k_B T),$$
(29)

with free energies defined in analogy with those in equations 25-27. Similarly, reactions equilibrating solutes on β1 and X sublattices include:

$$S_{\beta 1} \rightarrow S_X + V_{\beta 1}, \qquad K_{d1} = \frac{[S_X]}{[S_{\beta 1}]}[V_{\beta 1}] = \exp(-G_{d1}/k_B T); \qquad (30)$$

$$S_{\beta 1} + B_X \rightarrow S_X, \qquad K_{e1} = \frac{[S_X]}{[S_{\beta 1}]}[B_X]^{-1} = \exp(-G_{e1}/k_B T); \qquad (31)$$

$$S_{\beta 1} + A_X \rightarrow S_X + A_{\beta 1}, \qquad K_{f1} = \frac{[S_X]}{[S_{\beta 1}]}\frac{[A_{\beta 1}]}{[A_X]} = \exp(-G_{f1}/k_B T), \qquad (32)$$

$$S_{\beta 1} \rightarrow S_X + A_{\beta 1} + V_{\alpha}, \qquad K_{g1} = \frac{[S_X]}{[S_{\beta 1}]}[A_{\beta 1}][V_{\alpha}] = \exp(-G_{g1}/k_B T). \qquad (33)$$

leading to alternative expressions for the site-fraction ratio

$$R_X^{\beta 1} \equiv \frac{f_{\beta 1}}{f_X} = \frac{[S_{\beta 1}]}{[S_X]} = [V_{\beta 1}]\exp(+G_{d1}/k_B T) = [B_X]^{-1}\exp(+G_{e1}/k_B T) =$$
$$= \frac{[A_{\beta 1}]}{[A_X]}\exp(+G_{f1}/k_B T) = [A_{\beta 1}][V_{\alpha}]\exp(-G_{g1}/k_B T)$$
(34)

with a corresponding result for β2 and X sublattices:

$$R_X^{\beta 2} \equiv \frac{f_{\beta 2}}{f_X} = \frac{2[S_{\beta 2}]}{[S_X]} = 2[V_{\beta 2}]\exp(+G_{d2}/k_B T) = 2[B_X]^{-1}\exp(+G_{e2}/k_B T)$$
$$= \frac{2[A_{\beta 2}]}{[A_X]}\exp(+G_{f2}/k_B T) = 2[A_{\beta 2}][V_\alpha] = \exp(-G_{g2}/k_B T)$$
(35)

For $\alpha$ and $X$ sublattices, equilibrating reactions include:

$$S_\alpha \rightarrow S_X + V_\alpha, \qquad K_h = \frac{[S_X]}{[S_\alpha]}[V_\alpha] = \exp(-G_h/k_B T);$$ (36)

$$S_\alpha + A_X \rightarrow S_X, \qquad K_i = \frac{[S_X]}{[S_\alpha]}[A_X]^{-1} = \exp(-G_i/k_B T);$$ (37)

$$S_\alpha + A_\beta \rightarrow S_X + V_\beta, \qquad K_j = \frac{[S_X]}{[S_\alpha]}\frac{[V_\beta]}{[A_\beta]} = \exp(-G_j/k_B T),$$ (38)

from which

$$R_X^\alpha \equiv \frac{f_\alpha}{f_X} = \frac{2[S_\alpha]}{[S_X]} = 2[V_\alpha]\exp(+G_h/k_B T) = 2[A_X]^{-1}\exp(+G_i/k_B T) = 2\frac{[V_\beta]}{[A_\beta]}\exp(+G_j/k_B T). \quad (39)$$

Finally, equilibration of solutes on β1- and β2-sublattices takes place via simple exchange and is independent of intrinsic defect concentrations because only host B-atoms are involved, which have the same energy on both β-sublattices:

$$S_{\beta 2} + B_{\beta 1} \rightarrow S_{\beta 1} + B_{\beta 2}, \qquad K_k = \frac{[S_{\beta 1}]}{[S_{\beta 2}]} = \exp(-G_k/k_B T),$$ (40)

and

$$R_{\beta 1}^{\beta 2} \equiv \frac{f_{\beta 2}}{f_{\beta 1}} = \frac{2[S_{\beta 2}]}{[S_{\beta 1}]} = 2\exp(+G_k/k_B T).$$ (41)

Elsewhere, we shall show from PAC experiments on $Ni_2Al_3$ aluminides and galliumides that indium solutes in TM-rich samples strongly prefer the β2 sublattice, from which one can conclude that $G_k$ is positive.

Tab.VII

Selected expressions for $R$ are collected in Table VII, with definitions of activation energies for solute transfer given in terms of differences in solute site-energies and appropriate defect energies. If solute atoms are on regular sites of the structure, the site fractions must sum to unity:

$$f_\alpha + f_X + f_{\beta 1} + f_{\beta 2} = 1.$$ (42)

Individual site fractions can be calculated from a full set of site-fraction ratios. For example,

$$f_{\beta 2} = \frac{1}{1 + 1/R^{\beta 2}_{\beta 1} + 1/R^{\beta 2}_{\alpha} + 1/R^{\beta 2}_{X}}. \tag{43}$$

Fig. 8

Fig. 9

<u>Simulation of site-fraction ratios</u>. Site fraction ratios were simulated using defect concentrations previously simulated (see Figs. 6 and 7.) Energies of the solute atom on sites $\alpha$, X, $\beta 1$ and $\beta 2$ were chosen to be 0.75, 0.75, 0.75 and 0.30 eV, respectively. The energy of the solute on site $\beta 2$ was made lowest to accord with experimental results for In solutes in $Ni_2Al_3$ phases. Site-fraction ratios were calculated using expressions in Table VII and are plotted in Figure 8 as a function of the deviation from stoichiometry. It can be seen that pairs of site-fraction ratios such as $R^{\beta 2}_X$ and $R^{\beta 1}_X$, or $R^{\beta 2}_\alpha$ and $R^{\beta 1}_\alpha$, are offset from each other by the value $R^{\beta 2}_{\beta 1} = 65$, which is independent of composition. Presentation of the site-fractions themselves is more useful for comparison with measurement. In Fig. 9 are shown site-fractions calculated for using eq. 43 and analogous equations for the other fractions. $f_{\beta 2}$ is close to 100% for A-rich alloys but decreases in A-deficient alloys, being replaced by solutes on site $\alpha$ and to a lesser extent by solutes on site X.

The following observations were made for $Ni_2Al_3$ based on Figs. 8 and 9, other simulations and algebraic analysis:

1. As the composition deviates from stoichiometry, solutes relatively prefer the sublattice of the element in which the alloy is deficient. All things being equal, solutes have a propensity to sit on $\beta$-sublattices in B-poor alloys (x>0) and the $\alpha$-sublattice in A-poor alloys (x<0). In this way, the overall defect count is reduced.

2. The empty sublattice, X, has the character of an interstitial sublattice, with a maximum value of the site fraction near the stoichiometric composition (compare Fig. 5.)

3. Site preference is governed quantitatively by the relative energies of solutes on the different sites and by the energy of the relevant intrinsic defect(s). For the 8-defect model, the transfer energy $G_\alpha \equiv G(S_\alpha) - G(S_\beta) + G(A_\beta)$ must satisfy certain conditions. The conditions can be determined in the same way as for the CsCl structure in the preparation of Table V. Specifically, it can be shown that, at low temperature, (a) the solute will be on site $\alpha$ for all x if $G_a<0$, (b) the solute will switch from site $\alpha$ for x<0 to site $\beta$ for x>0 if $0<G_a<G_8/5$, and (c) the solute will remain on site $\beta$ for all x if $G_a>G_8/5$.

## VI.    DISCUSSION

<u>Comparison with other solute site-preference models</u>. Various features of the preference of solutes for substitutional sites have been obtained in other studies. In a study of Cu solutes in NiAl, Jacobi and Engell developed a thermodynamic model starting from the law of mass action while assuming that only triple-defects were present. The authors obtained S-shaped curves for the concentration of Cu solutes similar to those in the present study. This early study pointed out that the site preference of solutes depends strongly on the concentration of intrinsic defects and that a non-negligible concentration of solute affects site preferences. Later simulations have mostly dealt separately with the influences of temperature, composition or solute concentration on site preferences. In a study of Fe site preference in NiAl, Fu and Zou showed how site preferences changed for non-negligible solute concentrations.[21] In a study of solute site preference in $L1_2$ intermetallics[35] Wu et al. obtained a mapping of three types of solute site substitution behaviors with conditions based on pair interaction energies similar to those given in columns 1-3 in Table V. In addition, they showed that the width of the step-like discontinuity increases with increasing temperature and that a change in solute site preference may occur as a function of temperature (cf. Fig. 3 and discussion in the present work). Finally, they showed that the width of the step increases with increasing solute concentration.[35] Other studies have employed or developed similar mappings. An early study of solute incorporation in $L1_2$ phases correlated solute site-preference behaviors with the composition dependence of free energy curves near the stoichiometric composition.[36] There has been greater interest in determining how site-preference behavior depends on the type of solute in recent computational studies.[44, 45, 46]

Woodward et al. emphasized the importance of defect concentrations in determining site-preference behavior by developing analytical expressions for solute site-fraction ratios. In doing so, they made a stronger connection between defect thermodynamics and observed site-preference behavior than previous workers. The present model makes a similar connection while providing an analytical framework for investigating preference behavior of substitutional and interstitial solutes in a compound of arbitrary structure, stoichiometry, and defect type. Analytic forms for site-fraction ratios that have been obtained both by Woodward et al. and in the present paper are in complete agreement.

Site preferences in a 'line' compound. Frequently, phases in phase diagrams appear as a vertical line at the stoichiometric composition because the width of the phase field width has not been measured. However, at finite temperature such phases, called line compounds, must have a finite field width, no matter how small, and the composition range may not even encompass the stoichiometric composition. Because of the finite, perhaps unrecognized, width, routine sample preparations are then likely to lead to one of the two boundary compositions, with correspondingly different point defect concentrations and site preferences. If a solute in the phase tends to switch from one substitutional site to another as the composition changes, thefn, due to the unrecognized field width, the observed site preference may appear to vary randomly from sample to sample. A measured difference in site-preference in samples deliberately prepared to have the two boundary compositions would, using suitable energy parameters, allow one to estimate the width of a narrow phase field.

## VII. SUMMARY OF SITE PREFERENCE BEHAVIOR

The thermodynamic model provides clear predictions for solute site preferences based on general considerations of the analytic forms of expressions and on detailed simulations for CsCl and $Ni_2Al_3$ structures. The following trends have been found.

1. The preference of a solute between two sites is governed by site-energies of the solute, the formation energy of the dominant defect combination, and the site energy of an intrinsic defect (cf. Table V.)

2. Each site-fraction ratio is directly or inversely proportional to the concentration of an intrinsic defect or to a quotient or product of concentrations of two defects. Thus, for example, the site-fraction ratio $R_\alpha^\beta$ is proportional to $[A_\beta]$, $[V_\beta]/[V_\alpha]$, or $[B_\alpha]^{-1}$ in any binary structure and $R_\tau^\alpha$ is proportional to $[V_\alpha]$. Each ratio is also proportional to a Boltzmann factor containing a solute-transfer energy that involves the difference of energies of the solute on the two sites and the energy of the corresponding intrinsic defect or defects.

3. Solutes tend to occupy β-sites in a B-deficient alloy (in which $[A_\beta]$ or $[V_\beta]$ is large) and to occupy α-sites in an A-deficient alloy (in which $[V_\alpha]$ or $[B_\alpha]$ is large.) For a compound with a broad homogeneity range, step-like discontinuities in site-fraction ratios occur that are centered at the stoichiometric composition (cf. Figs. 2, 3 and 7.) The magnitude of the step is inversely proportional to the equilibrium constant for formation of the dominant defect combination. At low temperature, the range of composition over which a switch occurs may be very small (of the order of a percent.)

4. For a sufficiently large solute site-energy difference, the solute will always sit on the site where its energy is lowest (cf. columns 2 and 4 of Table V for substitutional sites.)

5. For intermediate values of the site-energy difference, the solute may switch from one site to another as the composition is changed (cf. column 3 of Table V and Figs. 4, 5 and 9 for substitutional sites; cf. Figs. 5 and 9 for interstitial sites) or as the temperature is changed (cf. Fig. 3.)

6. Since site fractions are determined from summations of ratios of arbitrary magnitude (cf. eq. 43) the composition at which a switch in site-fraction is half complete will not in general be precisely at the stoichiometric composition (cf. Fig. 4.) Whether or not a switch in site preference is observable depends on the phase boundary compositions, the width of the phase field, and the composition at which a switch is half complete.

7. Substitutional site fractions vary monotonically as a function of composition. A preference for interstitial sites is correlated with a low value for the sum of structural defect concentrations, and therefore the interstitial site fraction has a maximum value at or near the stoichiometric composition (cf. Fig. 5 and 9.)

8. The above phenomenological trends are found to be independent of the identity of the dominant defect combination and of the crystal structure.

ACKNOWLEDGMENTS

This work was supported in part by the National Science Foundation under grant DMR 00-91681 (Metals Program.)

APPENDIX:  EQUATIONS OF CONSTRAINT AMONG DEFECT CONCENTRATIONS.

Fundamental to the model are equations of constraint among defect concentrations. These are derived here for a binary compound of arbitrary stoichiometry and containing a non-negligible concentration of solute. It will be assumed that the unit cell contains a basis of $a$ substitutional A-type sites, $b$ substitutional B-type sites, and, in addition, $t$ sites of interstitial or empty-lattice type, all of which may be occupied by host or solute atoms. The sublattices corresponding to different types of sites will for the present be assumed to be inequivalent so that each will in general have different defect and solute site energies and, consequently, concentrations. Sublattices will be designated by subscripts $\alpha m, (m=1,a)$, $\beta n, (n=1,b)$ and $\tau k (k=1,t)$. Thus, there is a total of $a+b+t$ sites in the unit cell, of which $c \equiv a+b$ are substitutional and $t$ are interstitial. Let the total number of unit cells be $N$.

Total numbers of host atoms A and B and atoms of a particular solute species, S, will be assumed to be constant, with the composition determined by the formula $A_{a+cx}B_{b-cx}S_y$, in which x marks the deviation of the ratio of host elements from a stoichiometric composition $A_aB_b$ and y indicates the solute content. The total number of A-atoms is partitioned among the sublattices as

$$N_{tot}^{A} = \sum_{m=1}^{a} N_{\alpha m}^{A} + \sum_{n=1}^{b} N_{\beta n}^{A} + \sum_{k=1}^{t} N_{\tau k}^{A}, \tag{A1}$$

in which $N_{\alpha m}^{A}$ is the number of A-atoms on the $\alpha m$-th sublattice in all cells, *et cetera*, with expressions similar to eq. A1 for the numbers of B and S atoms. Eq. A1 is now rewritten in terms of fractional concentrations of atoms and defects, defined as the numbers of occupied sites divided by $N$ and written as symbols in square brackets; for example, the fractional concentration of vacancies on sublattice $\beta 2$ is $[V_{\beta 2}] \equiv N_{\beta 2}^{V}/N$. Eq. A1 becomes

$$N_{tot}^{A}/N = \sum_{m=1}^{a} [A_{\alpha m}] + \sum_{n=1}^{b} [A_{\beta n}] + \sum_{k=1}^{t} [A_{\tau k}]. \tag{A2}$$

The sums of atoms and defects on each sublattice equal $N$. For atoms and defects on the α sublattices, for example, this leads to $a$ relations

$$N = N_{\alpha m}^{A} + N_{\alpha m}^{B} + N_{\alpha m}^{S} + N_{\alpha m}^{V}, \quad (m=1,a), \tag{A3}$$

in which $V$ is a vacant site, and similarly for the $b$ $\beta$-sublattices and $t$ $\tau$-sublattices. Expressed in terms of atom and defect concentrations, eqs. A3 give

$$1 = [A_{\alpha m}] + [B_{\alpha m}] + [S_{\alpha m}] + [V_{\alpha m}], \quad (m=1,a), \tag{A4}$$

with similar expressions for $\beta$-sublattices and $\tau$-sublattices.

Two compositional constraints are imposed by the formula $A_{a+cx}B_{b-cx}S_y$. The first is a fixed ratio of numbers of A and B atoms:

$$\frac{N^A_{tot}}{N^B_{tot}} = \frac{a+cx}{b-cx}. \tag{A5}$$

The second is a fixed ratio of numbers of S and host atoms:

$$N^S_{tot} = \tfrac{y}{c}\left(N^A_{tot} + N^B_{tot}\right). \tag{A6}$$

Equations A5 and A6 can be rewritten using eq. A2 and corresponding equations for B and S atoms. In both equations, concentrations of host atoms on their normal sublattices that are normally large, such as $[A_{\alpha m}]$, are eliminated using eqs. A4 and similar sets of equations for $\beta$-ant $\tau$-sublattices. From eq. A5 one obtains

$$\sum_{m=1}^{a}[B_{\alpha m}] + (\tfrac{b}{c}-x)\sum_{m=1}^{a}\left([S_{\alpha m}]+[V_{\alpha m}]\right) + (\tfrac{a}{c}+x)\sum_{k=1}^{t}[B_{\tau k}] + cx =$$
$$= \sum_{n=1}^{b}[A_{\beta n}] + (\tfrac{a}{c}+x)\sum_{n=1}^{b}\left([S_{\beta n}]+[V_{\beta n}]\right) + (\tfrac{b}{c}-x)\sum_{k=1}^{t}[A_{\tau k}] \tag{A7}$$

and from eq. A6,

$$\sum_{m=1}^{a}[S_{\alpha m}] + \sum_{n=1}^{b}[S_{\beta n}] + \sum_{k=1}^{t}[S_{\tau k}] = \left(\frac{y}{y+c}\right)\left(c - \sum_{m=1}^{a}[V_{\alpha m}] - \sum_{n=1}^{b}[V_{\beta n}] + \sum_{k=1}^{t}\left([A_{\tau k}]+[B_{\tau k}]\right)\right). \tag{A8}$$

These two equations of constraint are valid for arbitrary crystal structures and composition, under the sole condition that defects are non-interacting. Below, we simplify eqs. A7-A9 for several practical situations and examine qualitative features of the equations of contraint.

(a) <u>Equivalent sublattices</u>. When sublattices are equivalent, they have equal defect concentrations and can be grouped together in eqs. A7 and A8. If $\alpha$-, $\beta$- and $\tau$-sublattices are all separately equivalent, then eq. A7 reduces to

$$a[B_\alpha] + a(\tfrac{b}{c}-x)\left([S_\alpha]+[V_\alpha]\right) + t(\tfrac{a}{c}+x)[B_\tau] + cx = b[A_\beta] + b(\tfrac{a}{c}+x)\left([S_\beta]+[V_\beta]\right) + t(\tfrac{b}{c}-x)[A_\tau], \tag{A9}$$

in which numerical subscripts of equivalent sublattices have been dropped. Concentrations of possible structural defects are located on the left- and right-hand sides of the equation for $x<0$ ($B_\alpha$, $V_\alpha$, $B_\tau$) and $x>0$ ($A_\beta$, $V_\beta$, $A_\tau$). Solutes can also behave as structural defects, reducing the overall defect count by appearing on the $\alpha$-sublattice in A-poor alloys ($x<0$) or $\beta$-sublattice in B-poor alloys, as long as the solute concentrations are less than the concentration of structural defects that would exist in the absence of solute. The absence of $[S_\tau]$ in eq. A9 is noteworthy because it signals that there is no direct coupling between the concentration $[S_\tau]$ and the composition, unlike for $[S_\alpha]$ and $[S_\beta]$. At stoichiometry, where there are no structural defects, the site preference of the solute will not reduce the defect count, so that--assuming nothing about defect and solute energies--solutes might equally well appear on either substitutional or interstitial sites. These same trends are found in other ways in the paper.

(b) <u>Negligible solute concentrations</u>. When solute concentrations can be neglected in comparison to concentrations of intrinsic defects, eq. A7 simplifies to

$$\sum_{m=1}^{a}[B_{\alpha m}] + (\tfrac{b}{c} - x)\sum_{m=1}^{a}[V_{\alpha m}] + (\tfrac{a}{c} + x)\sum_{k=1}^{t}[B_{\tau k}] + cx = \sum_{n=1}^{b}[A_{\beta n}] + (\tfrac{a}{c} + x)\sum_{n=1}^{b}[V_{\beta n}] + (\tfrac{b}{c} - x)\sum_{k=1}^{t}[A_{\tau k}]. \quad (A10)$$

Equations of constraint given for the CsCl and $Ni_2Al_3$ structures in eqs. 1 and 21, respectively, were obtained by inserting crystal structure information in eq. A10. The method is illustrated for the $Cu_3Au$ structure, which has three substitutional A-sublattices, all equivalent, and one B-sublattice. Let us also assume that there are two distinct types of interstitial sites of interest (for example one with octahedral symmetry and one with distorted octahedral symmetry. In eq. A10 one then has $a=3$, $b=1$, $c=a+b=4$, $t=2$, with the three A-sites equivalent and the two interstitial sites distinct. One then obtains

$$3[B_\alpha] + 3(\tfrac{1}{4} - x)[V_\alpha] + (\tfrac{3}{4} + x)([B_{\tau 1}] + [B_{\tau 2}]) + 4x = [A_\beta] + (\tfrac{3}{4} + x)[V_\beta] + (\tfrac{1}{4} - x)([A_{\tau 1}] + [A_{\tau 2}]), \quad (A11)$$

in which numerical subscripts for equivalent sublattices have been dropped.

(c) <u>Applications in the thermodynamic model</u>. Starting from the appropriate equation of constraint, the thermodynamic model is developed by reexpressing the equation in terms of a single defect concentration using expressions of equilibrium constants for (1) formation of defect combinations, such as in Tables I and VI, and (2) transfer of solute atoms among sublattices, such as given in eqs. 12-16 and analogous equations for $Ni_2Al_3$. Once one defect concentration has been determined for a given composition (x,y), temperature, and set of energy parameters, all others can be obtained using the formation reactions. In general, one has to consider distinct sublattices separately. When solute concentrations are not negligible, solute and defect concentrations are obtained explicitly using both eqs. A7 and A8. For example, expressed in terms of $[V_{\alpha 1}]$, in which $\alpha 1$ is an arbitrarily selected sublattice of A-type, it can be shown that the equation of constraint leads to a polynomial equation of degree 2c in $[V_{\alpha 1}]^{1/b}$ when solute concentrations are negligible and of degree 3c in $[V_{\alpha 1}]^{1/b}$ when they are not negligible.

Table I. Equilibrium defect combinations in the CsCl structure. Formation of a new unit cell is indicated by '+uc' in the reaction.

| Line | Equilibrium defect | Formation reaction | Equilibrium constant |
|---|---|---|---|
| 1 | Vacancy pair | $0 \to V_\alpha + V_\beta + uc$ | $K_{2V} = [V_\alpha][V_\beta] = \exp(-G_{2V}/k_B T)$ |
| 2 | Antisite pair | $0 \to A_\beta + B_\alpha$ | $K_{2A} = [A_\beta][B_\alpha] = \exp(-G_{2A}/k_B T)$ |
| 3 | Triple defect | $0 \to 2V_\alpha + A_\beta + uc$ | $K_3 = [V_\alpha]^2[A_\beta] = \exp(-G_3/k_B T)$ |
| 4 | Frenkel-A | $0 \to V_\alpha + A_\tau$ | $K_{2FA} = [V_\alpha][A_\tau] = \exp(-G_{2FA}/k_B T)$ |
| 5 | Frenkel-B | $0 \to V_\beta + B_\tau$ | $K_{2FB} = [V_\beta][B_\tau] = \exp(-G_{2FB}/k_B T)$ |

Table II. Defect concentrations in CsCl structures of composition $A_{1+2x}B_{1-2x}$ at stoichiometry and far from stoichiometry.

| Dominant defect | Defect | Defect concentration | | |
|---|---|---|---|---|
| | | x<<0 | x=0 | x>>0 |
| Antisite pair | $A_\beta$ | $K_{2A}/2|x|$ | $K_{2A}^{1/2}$ | $2x$ |
| Triple defect | $A_\beta$ | $K_3/16x^2$ | $K_3^{1/3}/4^{1/3}$ | $2x$ |
| Vacancy pair | $V_\alpha$ | $4|x|$ | $K_{2V}^{1/2}$ | $K_{2V}/4x$ |
| Frenkel-A | $V_\alpha$ | $4|x|$ | $\sqrt{3}K_{2FA}^{1/2}$ | $3K_{2FA}/4x$ |
| Frenkel-B | $V_\beta$ | $3K_{2FB}/4|x|$ | $\sqrt{3}K_{2FB}^{1/2}$ | $4x$ |

Table III. Expressions for site-fraction ratios for the CsCl structure and activation energies expressed in terms of defect and solute site-energies.

| Line | Site fraction ratio | Activation energy |
|---|---|---|
| 1 | $R_\alpha^\beta = [A_\beta]\exp(+G_a/k_B T)$ | $G_a = G(S_\alpha) - G(S_\beta) + G(A_\beta)$ |
| 2 | $R_\alpha^\beta = [B_\alpha]^{-1} \exp(+G_c/k_B T)$ | $G_c = G(S_\alpha) - G(S_\beta) - G(B_\alpha)$ |
| 3 | $R_\alpha^\beta = \dfrac{[V_\beta]}{[V_\alpha]}\exp(+G_b/k_B T)$ | $G_b = G(S_\alpha) - G(S_\beta) + G(V_\beta) - G(V_\alpha)$ |
| 4 | $R_\tau^\alpha = \tfrac{1}{3}[V_\alpha]\exp(+G_d/k_B T)$ | $G_d = G(S_\tau) - G(S_\alpha) + G(V_\alpha)$ |
| 5 | $R_\tau^\beta = \tfrac{1}{3}[V_\beta]\exp(+G_e/k_B T)$ | $G_e = G(S_\tau) - G(S_\beta) + G(V_\beta)$ |

Table IV. Expressions for site-fraction ratios in different composition regimes for CsCl structures of composition $A_{1+2x}B_{1-2x}$ that have a single dominant defect combination.

| Dominant defect combination | Site-fraction ratio | Expressions for specified compositions | | |
|---|---|---|---|---|
| | | $x \ll 0$ | $x = 0$ | $x \gg 0$ |
| Antisite pair | $R_\alpha^\beta$ | $K_{2A} K_a^{-1}/2|x|$ | $K_{2A}^{1/2} K_a^{-1}$ | $2x K_a^{-1}$ |
| Triple defect | $R_\alpha^\beta$ | $K_3 K_a^{-1}/16x^2$ | $K_3^{1/3} K_a^{-1}/4^{1/3}$ | $2x K_a^{-1}$ |
| Vacancy pair | $R_\alpha^\beta$ | $K_{2V} K_b^{-1}/16x^2$ | $K_b^{-1}$ | $16x^2 K_{2V}^{-1} K_b^{-1}$ |
| Frenkel-A | $R_\tau^\alpha$ | $4|x| K_d^{-1}/3$ | $K_{2FA}^{1/2} K_d^{-1}/\sqrt{3}$ | $K_{2FA} K_d^{-1}/4x$ |
| Frenkel-B | $R_\tau^\beta$ | $K_{2FB} K_e^{-1}/4|x|$ | $K_{2FB}^{1/2} K_e^{-1}/\sqrt{3}$ | $4x K_e^{-1}/3$ |

Table V. Predicted dependence of substitutional solute site fractions in CsCl phases $A_{1+2x}B_{1-2x}$ on the composition. $\Delta G \equiv G(S_\alpha) - G(S_\beta)$ is defined as the difference of site-energies of solutes on the $\alpha$ and $\beta$ sublattices. Concentrations of host and solute interstitials are assumed to be negligible.

| Dominant defect | $f_\alpha \gg f_\beta$ for all x | $f_\alpha \sim 1$ for x<0; $f_\beta \sim 1$ for x>0 | $f_\beta \gg f_\alpha$ for all x | $f_\alpha \cong f_\beta$ at stoichiometry |
|---|---|---|---|---|
| Antisite pair | $\Delta G + G(A_\beta) < 0$ | $0 < \Delta G + G(A_\beta) < G_{2A}$ | $\Delta G + G(A_\beta) > G_{2A}$ | $\Delta G + G(A_\beta) = \dfrac{G_{2A}}{2}$ |
| Triple defect | $\Delta G + G(A_\beta) < 0$ | $0 < \Delta G + G(A_\beta) < G_3$ | $\Delta G + G(A_\beta) > G_3$ | $\Delta G + G(A_\beta) = \dfrac{G_3}{3}$ |
| Vacancy pair | $\Delta G + 2G(V_\beta) < 0$ | $0 < \Delta G + 2G(V_\beta) < 2G_{2V}$ | $\Delta G + 2G(V_\beta) > 2G_{2V}$ | $\Delta G + 2G(V_\beta) = G_{2V}$ |

Table VI. Equilibrium defects in the $Ni_2Al_3$ structure. The $\beta$ subscripts refer to either $\beta 1$ and $\beta 2$ sublattices, considered indistinguishable, or solely to sublattice $\beta 1$ if they are distinguishable. Creation or destruction of a unit cell in the course of defect formation is indicated by +uc and -uc, respectively.

| Line | Equilibrium defect | Formation reaction | Equilibrium constant |
|---|---|---|---|
| 1 | 5-vacancy | $0 \to 2V_\alpha + 3V_\beta + uc$ | $K_{5V} = [V_\alpha]^2[V_\beta]^3 = \exp(-G_{5V}/k_BT)$ |
| 2 | Antisite pair | $0 \to B_\alpha + A_\beta$ | $K_{2A} = [A_\beta][B_\alpha] = \exp(-G_{2A}/k_BT)$ |
| 3 | 8-defect | $0 \to 5V_\alpha + 3A_\beta + uc$ | $K_8 = [V_\alpha]^5[A_\beta]^3 = \exp(-G_8/k_BT)$ |
| 4 | Frenkel-A | $0 \to V_\alpha + A_x$ | $K_{2FA} = [V_\alpha][A_X] = \exp(-G_{2FA}/k_BT)$ |
| 5 | Frenkel-B | $0 \to V_\beta + B_x$ | $K_{2FB} = [V_\beta][B_X] = \exp(-G_{2FB}/k_BT)$ |
| 6 | Interstitial | $0 \to 2A_x + 3B_x - uc$ | $K_{5I} = [A_X]^2[B_X]^3 = \exp(-G_{5I}/k_BT)$ |
| 7 | 7-defect | $0 \to 5V_\beta + 2B_\alpha + uc$ | $K_7 = [V_\beta]^5[B_\alpha]^2 = \exp(-G_7/k_BT)$ |
| 8 | Anti-8-defect | $0 \to 5A_X + 3B_\alpha - uc$ | $K_{8X} = [A_X]^5[B_\alpha]^3 = \exp(-G_{8X}/k_BT)$ |
| 9 | Anti-7-defect | $0 \to 5B_X + 2A_\beta - uc$ | $K_{7X} = [B_X]^5[A_\beta]^2 = \exp(-G_{7X}/k_BT)$ |
| 10 | 3-defect | $0 \to B_\alpha + V_\beta + A_X$ | $K_3 = [B_\alpha][V_\beta][A_X] = \exp(-G_3/k_BT)$ |
| 11 | Anti-3-defect | $0 \to A_\beta + V_\alpha + B_X$ | $K_{3A} = [A_\beta][V_\alpha][B_X] = \exp(-G_{3A}/k_BT)$ |

Table VII. Selected expressions for site-fraction ratios of solutes in the $Ni_2Al_3$ structure.

| Line | Site-fraction ratio $R_{site2}^{site1} \equiv f_{site1}/f_{site2}$ | Activation energy in terms of site energies |
|---|---|---|
| 1 | $R_\alpha^{\beta 1} = \frac{1}{2}[A_{\beta 1}]\exp(+G_{a1}/k_B T)$ | $G_{a1} = G(S_\alpha) - G(S_{\beta 1}) + G(A_{\beta 1})$ |
| 2 | $R_\alpha^{\beta 2} = [A_{\beta 1}]\exp(+G_{a2}/k_B T)$ | $G_{a2} = G(S_\alpha) - G(S_{\beta 2}) + G(A_{\beta 1})$ |
| 3 | $R_X^{\beta 1} = [A_{\beta 1}][V_\alpha]\exp(+G_{g1}/k_B T)$ | $G_{g1} = G(S_X) - G(S_{\beta 1}) + G(A_{\beta 1}) + G(V_\alpha)$ |
| 4 | $R_X^{\beta 2} = 2[A_{\beta 2}][V_\alpha]\exp(+G_{g2}/k_B T)$ | $G_{g2} = G(S_X) - G(S_{\beta 2}) + G(A_{\beta 2}) + G(V_\varepsilon)$ |
| 5 | $R_X^\alpha = 2[V_\alpha]\exp(+G_h/k_B T)$ | $G_h = G(S_X) - G(S_\alpha) + G(V_\alpha)$ |
| 6 | $R_{\beta 1}^{\beta 2} = 2\exp(+G_k/k_B T)$ | $G_k = G(S_{\beta 1}) - G(S_{\beta 2})$ |

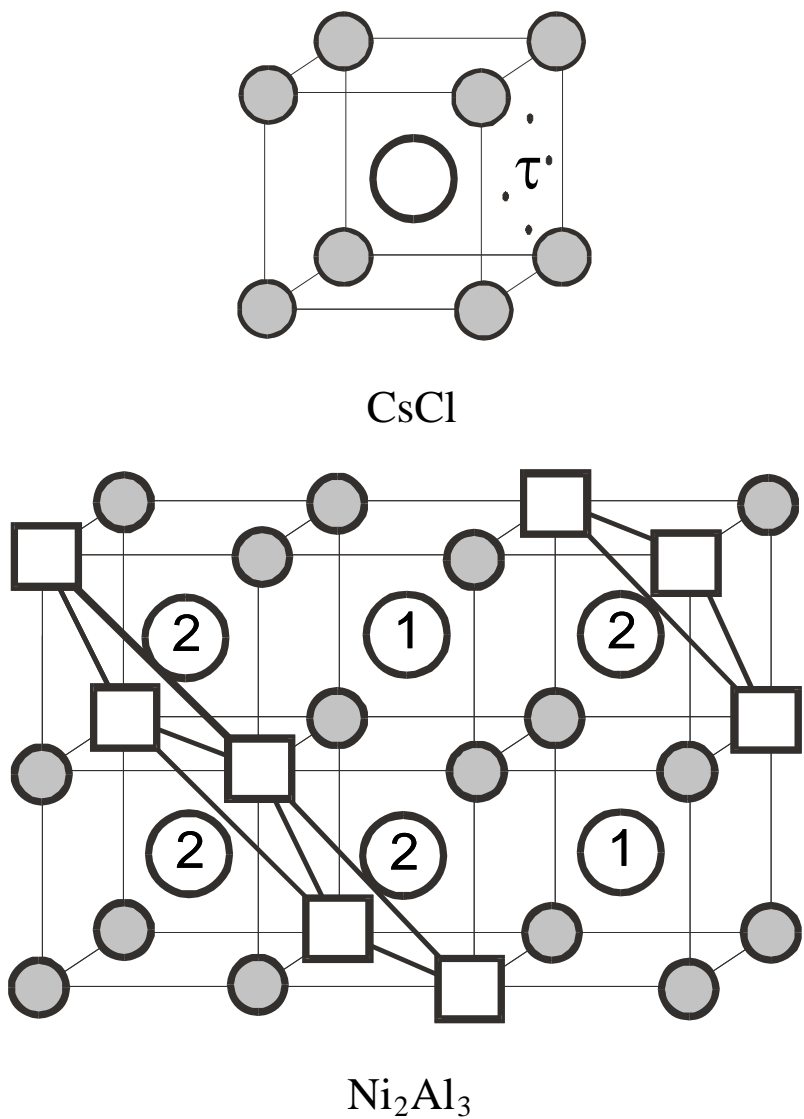

Figure 1. Crystal structures of CsCl and $Ni_2Al_3$. Atoms on $\alpha$- and $\beta$-sublattices are shown by small shaded circles and large open circles. For CsCl, distorted tetrahedral interstitial sites $\tau$ are also shown. For $Ni_2Al_3$, an empty sublattice is shown by squares. The actual $Ni_2Al_3$ structure is distorted slightly from the cubic arrangement shown. Numbers identify two inequivalent $\beta$-sites in the $Ni_2Al_3$ structure present in a ratio of 2:1.

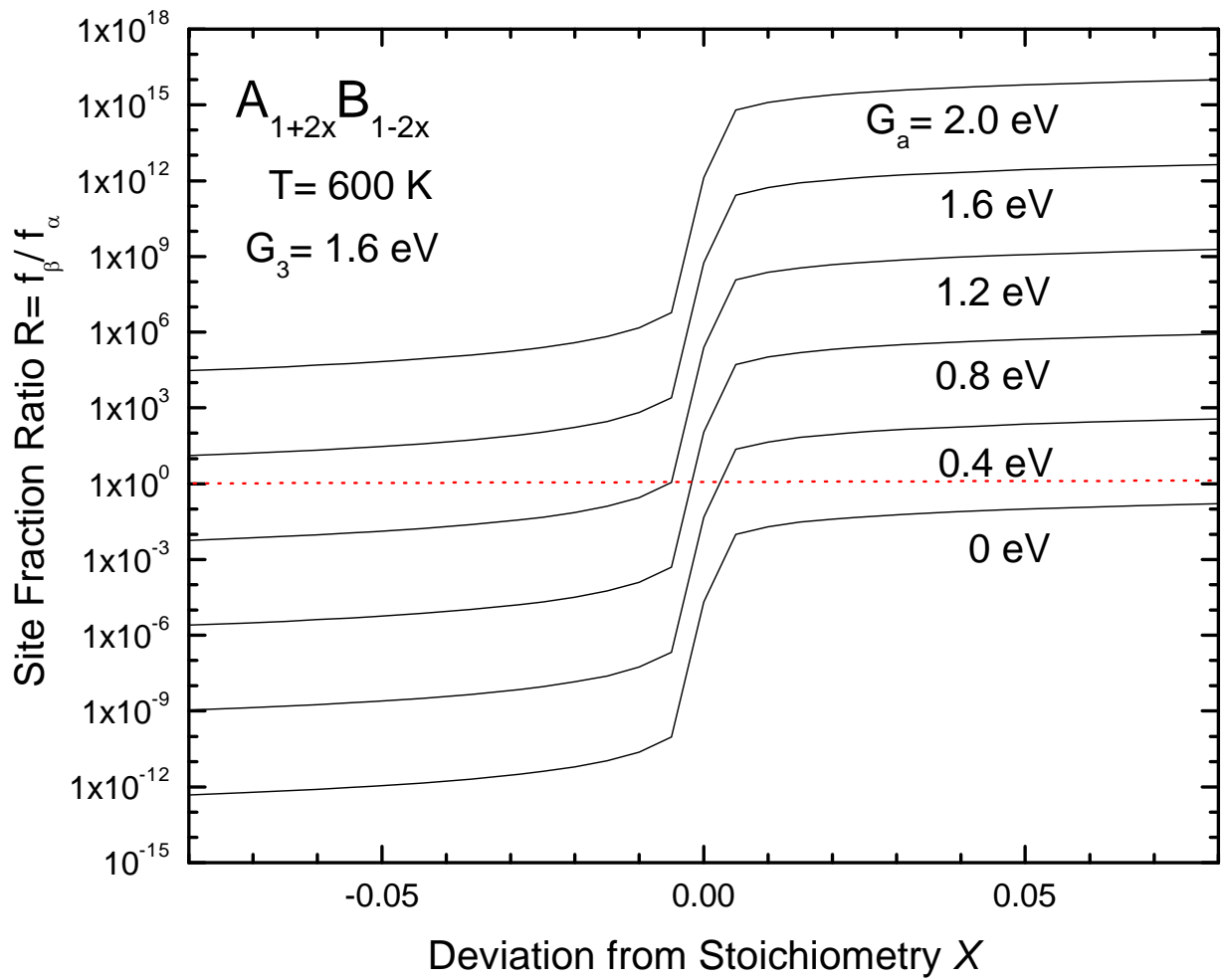

Figure 2. Site fraction ratio $R_\beta^\alpha$ at 600 K as a function of composition assuming the triple defect is dominant with formation energy $G_3 = 1.6$ eV and for various indicated values of the solute-transfer activation energy $G_\alpha$ from eq. 12. Unity ratio is indicated by the horizontal dashed line.

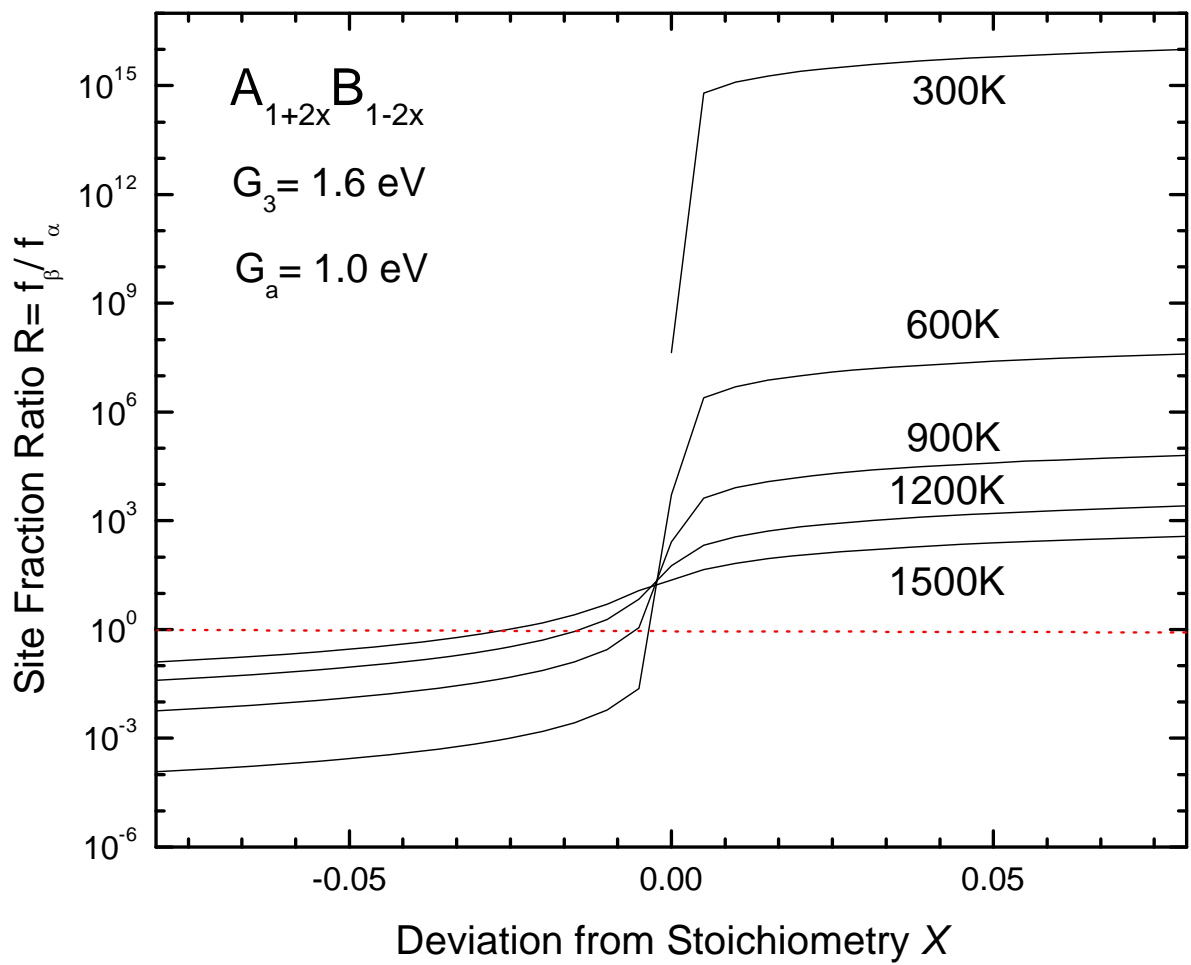

Figure 3. Site-fraction ratio $R_\beta^\alpha$ as a function of composition at the indicated temperatures. The formation energy of a triple-defect was fixed to 1.6 eV and the solute-transfer activation energy was fixed to 1.0 eV. Unity ratio is indicated by the horizontal dashed line.

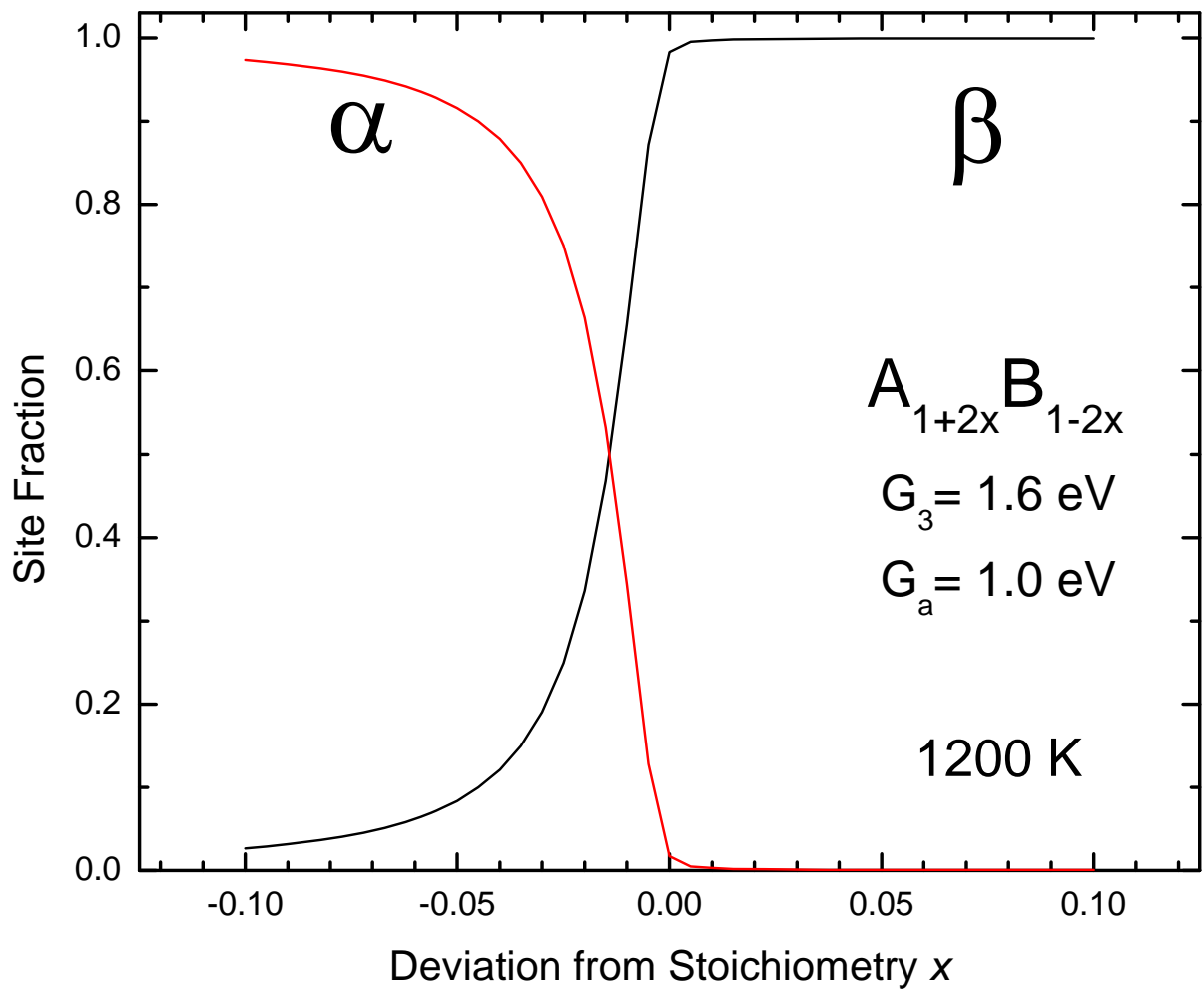

Figure 4. Site fractions of a solute on the α and β sublattices in the CsCl structure, calculated from the site-fraction ratio curve for 1200 K in Fig. 3 under the assumption that only α and β sites are occupied. The solute is observed to change site preference from the α-site for A-deficient compositions to the β-site for A-rich compositions.

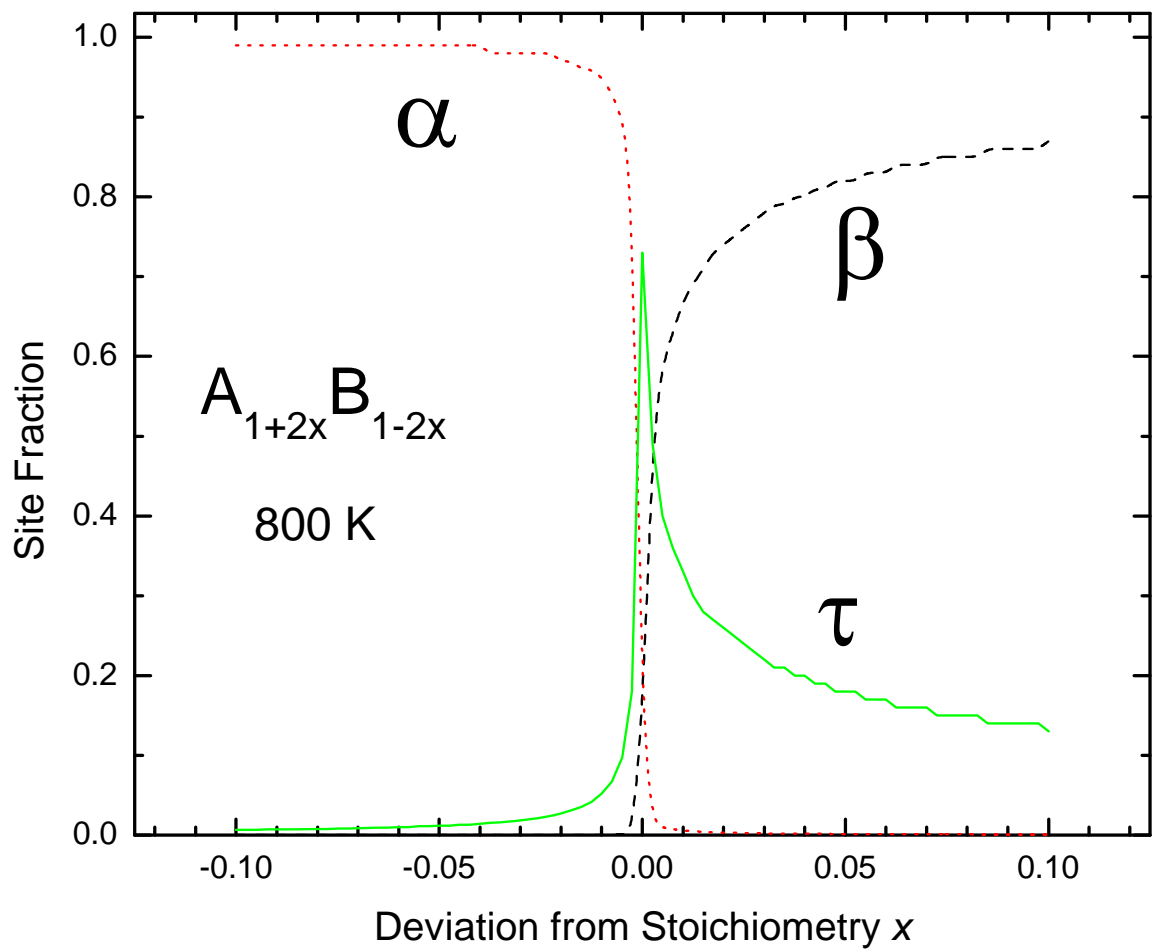

Figure 5. Site fractions of a solute on $\alpha$, $\beta$ and $\tau$ sublattices in CsCl. The solute changes preference from the $\alpha$-site for A-deficient compositions to the $\beta$-site for A-rich compositions, with a site-fraction on the $\tau$-sublattice peaking near the stoichiometric composition.

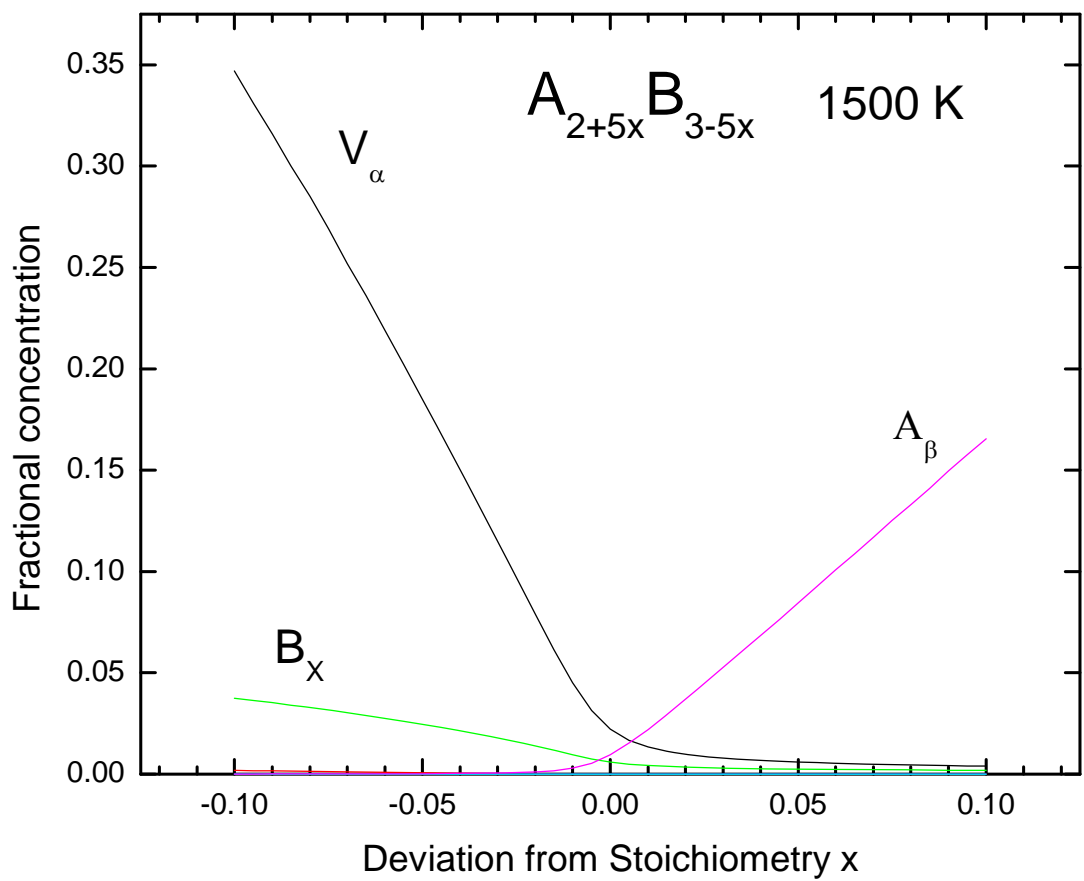

Figure 6. Fractional concentrations of elementary defects in $Ni_{2+5x}Al_{3-5x}$ as a function of composition. For the choice of model energies used, the structural defects are $V_A$ and $A_B$, constituents of the 8-defect ($5V_A + 3A_B$). Those defects are also most easily thermally activated, although a thermally activated concentration of $B_X$ is also observed.

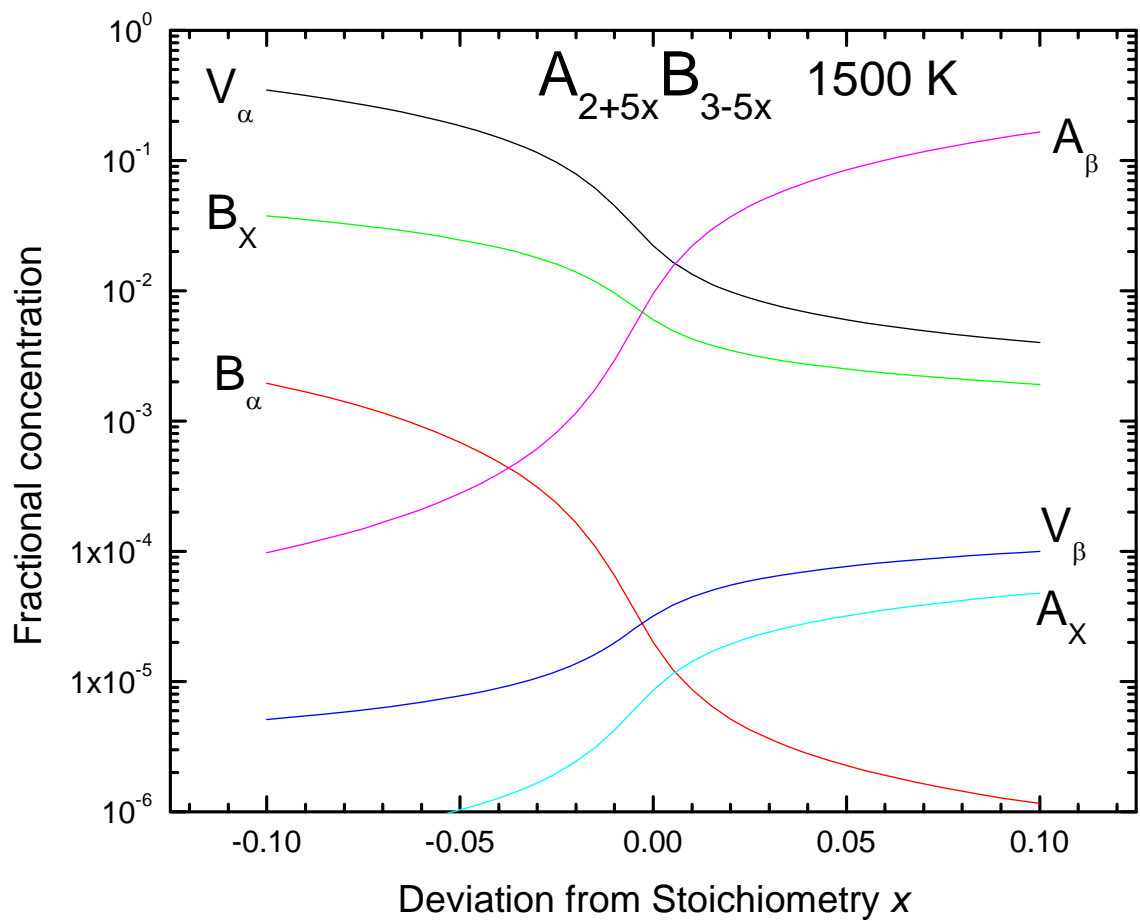

Figure 7. Log-plot of fractional concentration of elementary defects in $A_{2+5x}B_{3-5x}$ versus x. Same data as in Fig. 6, showing minor defect concentrations.

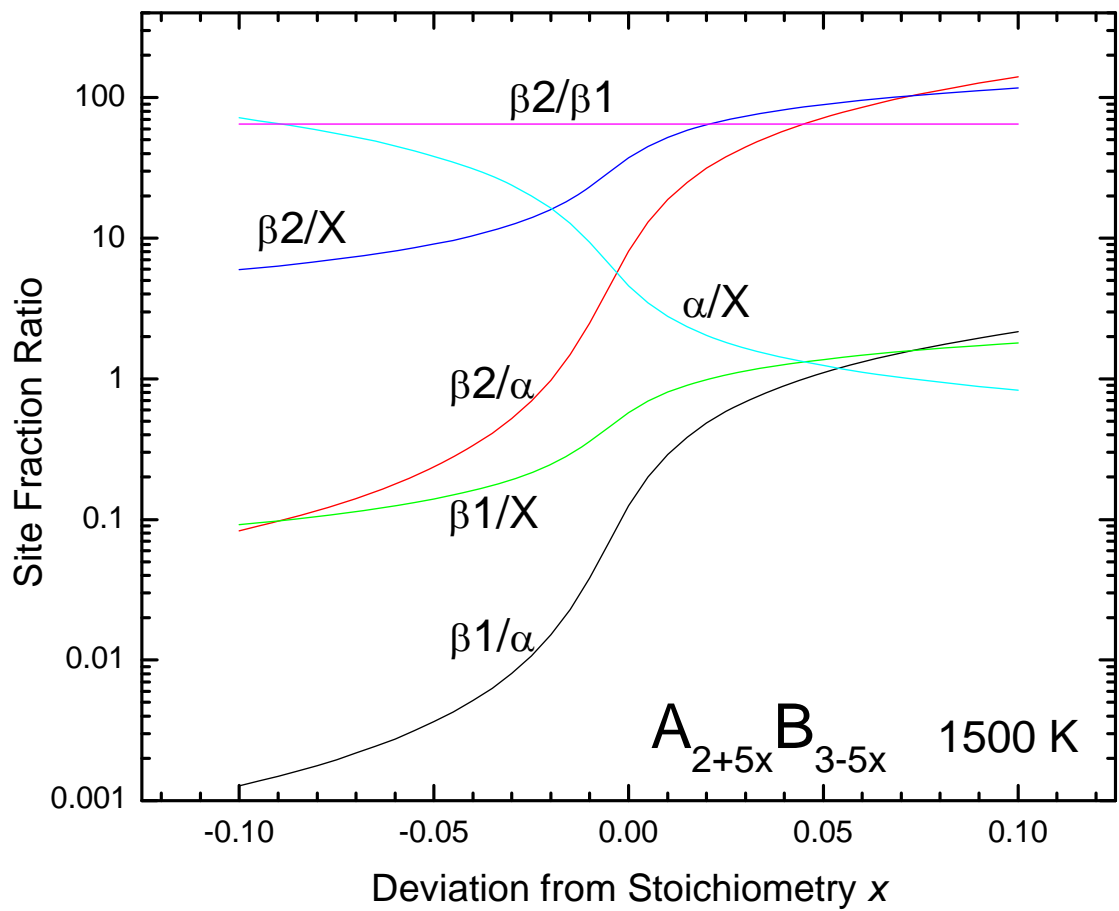

Figure 8. Ratios of fractions of solutes on different sites in $Ni_{2+5x}Al_{3-5x}$ calculated for defect concentrations shown in Fig. 6 and for site-energies specified in the text. Abbreviations identify site-fraction ratios; for example β2/α indicates $R_\alpha^{\beta 2} = f_{\beta 2}/f_\alpha$.

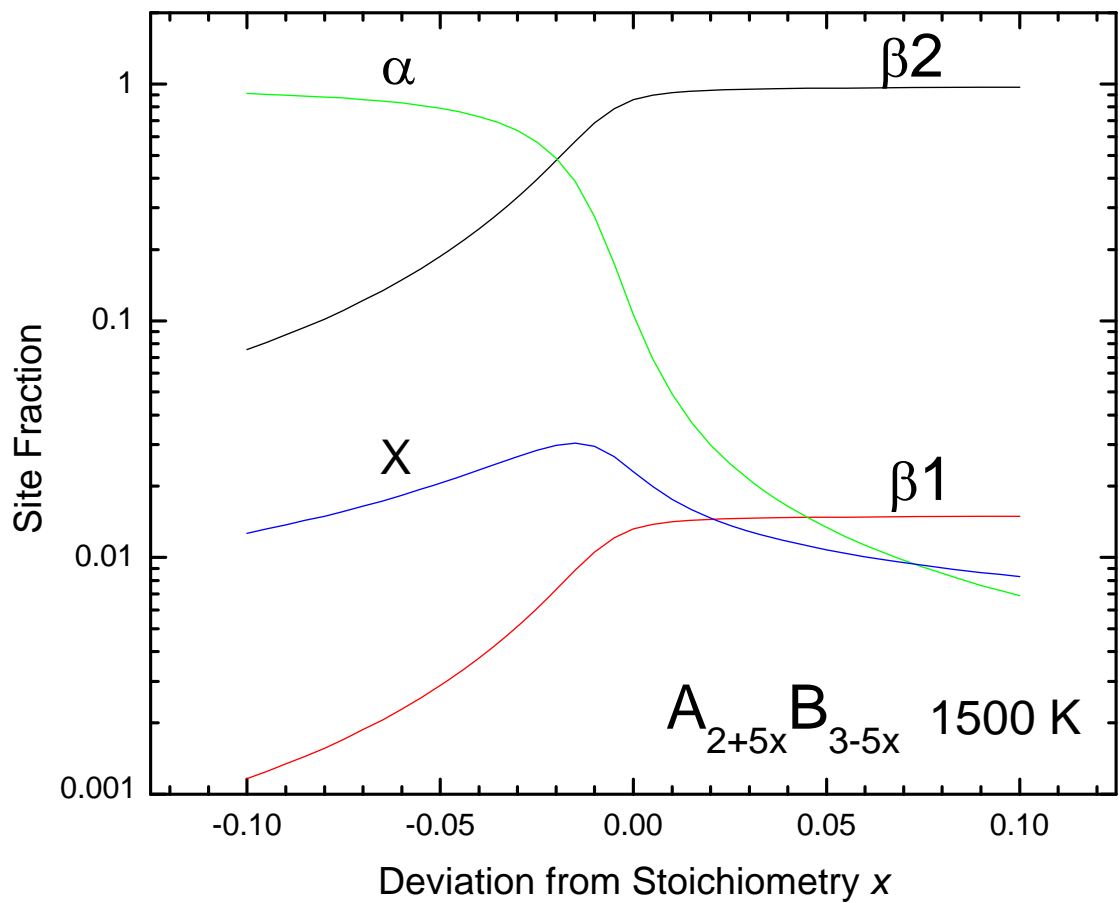

Figure 9. Log-plot of site fractions of solutes in $Ni_{2+5x}Al_{3-5x}$. Solutes are predominantly on A-sites for A-deficient compositions (x<0) and on B-sites for B-deficient compositions (x>0). A significant fraction of solutes occupy empty-lattice X-sites near the stoichiometric composition.